\documentclass[english]{article}
\usepackage[T1]{fontenc}
\usepackage[latin9]{inputenc}
\usepackage{geometry}
\usepackage{mdframed}
\usepackage{color}
\usepackage{graphicx}
\usepackage{verbatim}
\usepackage{amssymb}
\usepackage{epstopdf}
\usepackage{amsmath}
\usepackage{stmaryrd}
\usepackage{mathtools}
\usepackage{caption}
\usepackage[makeroom]{cancel}
\usepackage{indentfirst}
\usepackage{multirow}

\newcommand{\comments}[1]{}   

\makeatletter

\usepackage{indentfirst}
\usepackage{enumerate}
\usepackage{bigints}
\usepackage{cancel}
\usepackage{amsmath}
\usepackage[normalem]{ulem}

\definecolor{mygreen}{rgb}{.0,.5,.1}

\def\XXint#1#2#3{{\setbox0=\hbox{$#1{#2#3}{\int}$}
     \vcenter{\hbox{$#2#3$}}\kern-.5\wd0}}

\title{Finite size effects and loss of self-averageness in the relaxational dynamics of the spherical Sherrington-Kirkpatrick model}
\author{Damien Barbier$^1$, Pedro H. de Freitas Pimenta$^2$, Leticia F. Cugliandolo$^{1,3}$ and Daniel A. Stariolo$^2$\\
{\small 
$^1$Sorbonne Universit\'e, CNRS UMR 7589, Laboratoire de Physique Th\'eorique et Hautes Energies,}
\\
{\small 4 Place Jussieu, 75252 Paris Cedex 05, France}
\\
{\small $^2$Universidade  Federal  Fluminense,  Departamento  de  F\'{\i}sica  and}  \\
{\small National  Institute  of Science  and  Technology  for  Complex  Systems, } \\
{\small Av.   Gal.   Milton  Tavares  de  Souza  s/n, Campus  da  Praia  Vermelha,  24210-346  Niter\'oi,  RJ,  Brazil}
\\
{\small $^3$Institut Universitaire de France, 1, rue Descartes, 75231 Paris Cedex 05, France}
}

\makeatother

\begin{document}

\maketitle

\abstract{
We revisit the gradient descent dynamics of the spherical Sherrington-Kirkpatrick ($p=2$) model with finite number 
of degrees of freedom. For fully random initial conditions we confirm that
the relaxation takes place in three time regimes: a first algebraic one controlled by the decay of the eigenvalue distribution of 
the random exchange interaction matrix at its edge in the infinite size limit; a faster algebraic one determined by the distribution of the 
gap between the two extreme eigenvalues; and a final exponential one determined by the minimal gap sampled in the 
disorder average. We also analyse the finite size effects on the relaxation from initial states which are almost projected on 
the saddles of the potential energy landscape, and we show that for deviations scaling as $N^{-\nu}$ from perfect alignment 
the system escapes the initial configuration in a time-scale scaling as $\ln N$ after which the dynamics no longer ``self-averages''
with respect to the initial conditions. We prove these statements with a combination of analytic and numerical methods.}


\newpage

\tableofcontents

\newpage

\section{Introduction}

Fully connected or large dimensional models with quenched random interactions are  commonly used as mean-field models 
for physical systems. In addition, they are realistic models for problems in other branches of 
sciences. Examples are manifold and only some celebrated ones are neural networks~\cite{Hopfield82}, 
ecosystems~\cite{May72} and macro-economy agent-based models~\cite{Gualdi15}. In these areas, the 
number of degrees of freedom is large but much smaller than the Avogadro number, or the dimension of space is large but also finite. 
Therefore,  the ideal physical thermodynamical limit is 
a long way from being attained. Finite size or finite dimensional fluctuations can be very important and in 
some cases completely dominate the dynamic behaviour.

While a rather complete understanding of the asymptotic dynamics of disordered mean-field like models
in the strict thermodynamic limit has been achieved in the last decades~\cite{Cugliandolo03,BerthierBiroli,Zamponi}, the knowledge  of 
finite size fluctuations is not as developed. Having said so,  after a few studies in the early 2000s~\cite{CrRi00a,CrRi00c,CrRi04,Billoire2005}, 
there has been a resurgence of interest in the dynamics of finite size strongly interacting models with quenched 
randomness. These studies were boosted by the will to 
better grasp the relaxation of glassy systems. Most importantly, the goals were to distinguish the relaxation akin to gradient descent 
from activation over free-energy barriers and, especially, to quantify the role played by the latter in the approach to equilibrium~\cite{BaityJesi18a,BaityJesi18b,StCu19,StCu20,Ros20,BaityJesi21} (see also~\cite{BenArous18,Jagannath19,Dembo20} from the 
mathematics literature).

In this paper we picked the simplest disordered model we know of, the so-called spherical  $p=2$ or spherical Sherrington-Kirkpatrick
model~\cite{KoThJo76},
and we studied its relaxation at zero temperature in instances with a finite numbers of degrees of freedom or modes.
Around a decade ago, some studies focused on the finite size effects in equilibrium and the fluctuations induced by them~\cite{CuDeYo07,MoGa13,FyLeDou14}, 
and many more recent mathematics papers also deal with these~\cite{Baik16,Baik18,Kivimae19,Nguyen19,Landon19,Landon20,LeDou20}. Here we concentrate on the 
relaxation dynamics and we build upon previous 
works performed in the strict thermodynamic limit~\cite{ShSi81,CidePa88,CuDe95a,CuDe95b,BenArous01,ChCuYo06,vanDuJavanWij10} 
and for finite size systems~\cite{FyPeSc15}.
Concretely,
\begin{itemize}
\item[-]
for high-$T$ initial conditions we revisit the fluctuations with respect to the 
random matrix realisation,
\item[-]
for low-$T$ initial conditions - correlated with the random interaction matrix - we analyse the fluctuations with respect to the 
initial condition.
\end{itemize}

The paper is organised as follows. In Sec.~\ref{sec:model} we introduce the model and we recall, very 
briefly, a few of its properties. Section~\ref{sec:flat} is devoted to the analysis of the fluctuations 
induced by the random matrix realisation after sub-critical quenches from random (flat) initial conditions.
The next Sec.~\ref{sec:correlated} treats the case of initial conditions correlated with the random matrix. Finally,
in Sec.~\ref{sec:conclusions} we present our conclusions.

\section{The model}
\label{sec:model}

The spherical Sherrington-Kirkpatrick or $p=2$ disordered system is a model of $N$ pair-wise 
interacting ``spins'' taking real values~\cite{KoThJo76}. 
It is described by the Hamiltonian
\begin{equation}
    H[\vec{S},z]=-\frac{1}{2}\sum_{i\neq j}J_{ij}s_i s_j+\frac{z}{2} \left(\sum_i s_i^2-N\right) 
    =
    -\frac{1}{2}\vec{S} \cdot (\mathbf{J}\vec{S})+\frac{z}{2} \left(\vec{S}^{\, 2}-N\right)
    \; . 
    \label{eq:p2model}
\end{equation}
In the last expression the spins have been arranged in an $N$ component vector $\vec{S}=(s_1,\dots,s_N)$. 
The coupling constants $J_{ij}$ are real and symmetric, and form a matrix in the Gaussian
Orthogonal Ensemble (GOE), $\mathbf J=\{J_{ij}\}_{(i,j)\in [\![1,N]\!]²}$. They 
have mean and variance ${\rm I\!E} [J_{ij}]=0$ and ${\rm I\!E} [J_{ij}^2]=J^2/N$, the latter ensuring
extensivity of the energy in the thermodynamic limit. The energy of the model is given by just the first term in eq.~(\ref{eq:p2model})
and $z$ is a Lagrange multiplier enforcing the spherical constraint $|\vec S|^2=N$ which also reads
\begin{equation}
\vec S^{\, 2} =  \sum_i s_i^2=N 
\label{eq:spherical-constraint}
\; . 
\end{equation}
Straightforwardly, the Hamiltonian is simplified when $\vec{S}$ is decomposed using the orthonormal eigen-basis 
$\{\vec{V}_\mu\}$, with $\vec{V}_\mu \cdot \vec V_\nu= \delta_{\mu\nu}$, of the coupling matrix $\mathbf J$. With the notation 
\begin{equation}
s_\mu=\vec{S} \cdot \vec{V}_\mu
\end{equation}
for the projections of $\vec S$ in these directions, the Hamiltonian becomes
\begin{equation}
    H[\vec{S},z]=-\frac{1}{2}\sum_{\mu}(\lambda_\mu-z) s_\mu^2 -\frac{z}{2} \, N\, ,
\end{equation}
where $\{\lambda_\mu\}_{\mu \in [\![1,N]\!]}$ are the set of eigenvalues of $\mathbf J$ (with associated eigenvectors $\{\vec{V}_\mu\}$)
ordered such that $\lambda_N>\dots>\lambda_1$. 

In the infinite $N$ limit, 
the eigenvalues of GOE matrices are distributed according to the Wigner semi-circle law of radius $2J$
\begin{equation}
\rho(\lambda) = \frac{1}{2\pi J^2} \sqrt{(2J)^2 - \lambda^2} \qquad\qquad \mbox{for} \;\; \lambda \in [-2J, 2J] 
\end{equation}
and zero otherwise~\cite{Me91}. At high temperatures, $T>J$, the equilibrium state is paramagnetic. 
At $T=J$ there is a second order phase transition towards a low-temperature phase with two equilibrium states related by spin inversion 
symmetry~\cite{KoThJo76,FyLeDou14}, which makes this model
  closer to a disordered ferromagnet than to a true spin glass. Here and in what follows we set the Boltzmann constant to one.

  In spite of the relative simplicity of its
  thermodynamic structure, the out of equilibrium dynamics show many complex
features like slow relaxation and aging~\cite{ShSi81,CidePa88,CuDe95a,CuDe95b,BenArous01,ChCuYo06,vanDuJavanWij10,FyPeSc15}. 
  The damped dynamics are governed by the set of Langevin equations:
\begin{equation}
    \partial_t s_i(t)= \sum_{j(\neq i)} J_{ij} s_j(t)  - z\big(t,\{s_\mu(0)\}\big) s_i(t) + \xi_i(t) \quad\qquad \forall i \in [\![1,N]\!]
    \; ,
    \label{eq:langevin}
\end{equation}
where $\xi_i(t)$ represents a Gaussian white noise with zero mean and variance
$\langle \xi_i(t)\xi_j(t') \rangle = 2T\,\delta_{ij}\delta(t-t')$, and $T$ is the temperature of a thermal bath. 
We absorbed the friction coefficient in the definition of time. Accordingly,  the microscopic time-scale is 
in the new units equal to one.
 In the present work, we are interested in the deterministic limit of  eqs.~(\ref{eq:langevin}),
which in the basis of eigenvectors read
\begin{equation}
    \partial_t s_\mu(t)=\Big[\lambda_\mu-z\big(t,\{s_\mu(0)\}\big)\Big]s_\mu(t) \quad\qquad \forall\mu\in [\![1,N]\!]
    \; .
    \label{eq:langevin-mu}
\end{equation}
These dynamics can be considered to be the zero temperature limit of the Langevin equations~(\ref{eq:langevin}).
The initial conditions are $\{s_\mu(0)\}$ and in the expressions
 above we made explicit the fact that $z$ depends on time and on them as well.
 
In terms of thermodynamics the zero temperature free energy is directly its Hamiltonian $H[\vec{S},z]$. 
In the long-time limit the system must fall in a stable or metastable  state of the free energy, and 
which of them should depend on the initial condition.  By setting $\lim_{t\to\infty} \partial_t s_\mu(t) =0$ we obtain the criteria
\begin{equation}
    \delta_{s_\mu}H[\vec{S},z]=-(\lambda_\mu-z)s_\mu=0 \quad\qquad \forall \mu \in [\![1,N]\!]
\end{equation}
and at all times, complemented by the spherical constraint in eq.~(\ref{eq:spherical-constraint}).
The solutions to this system of equations  are 
\begin{equation}
\vec{S}=\pm\sqrt{N} \,  \vec{V}_\mu \qquad\qquad \mbox{and}  \qquad\qquad z=\lambda_\mu \quad\qquad \forall \mu \in [\![1,N]\!]
\; , 
\end{equation} 
and they are $2N$ in number.
Their stability is determined by the Hessian $\delta_{s_\mu}\delta_{s_\nu} H[\vec{S},z]=-\delta_{\mu,\nu}(\lambda_\mu-z)$. 
Taking a given metastable state $\vec{S}=\sqrt{N}\vec{V}_\mu$, the local landscape has $N-\mu$ stable directions, $\mu-1$ unstable directions and a marginal flat one. 
The energy of each of these configurations, or equivalently $H[\vec S=\sqrt{N} \vec V_\mu, z=\lambda_\mu]$ is simply equal to $- \lambda_\mu N/2$.
This energy landscape analysis predicts that the system should always equilibrate in one of the solutions 
$\pm\sqrt{N}\vec{V}_N$ as they are the only stable ones with respect to $H[\vec{S},z]$. The ground state energy density is then $e_{\rm eq} = - \lambda_N/2$.
We recall that $\lambda_N =2J + J \zeta N^{-2/3}$ with $\zeta$ a random variable distributed by the Tracy-Widom form~\cite{TrWi1,TrWi2}. 
Therefore, the equilibrium $z$ also depends 
on $N$ but we do not write it down explicitly to lighten the notation.

\subsection{Sources of fluctuations}

We are interested in the dynamic fluctuations induced by the finite system size. We consider two kinds of initial conditions, flat and projected (called staggered in~\cite{CuDe95a}), 
that we 
define in Sec.~\ref{subsec:initial}. In the former case, the fluctuations are induced by the random matrix  or, equivalently, by the set 
of eigenvalues $\{\lambda_\mu\}$,  which vary close to the edge of $\rho(\lambda)$ in a way that we recall in Sec.~\ref{subsec:edge}. 
In the latter case,  we focus on initial states that are strongly correlated with the random matrix, in the form of small 
deviations from perfect alignment with a metastable configuration $\vec V_\alpha$. 

\subsubsection{The initial conditions}
\label{subsec:initial}

The crucial point of the following study will be the choice of the initial condition $\vec{S}(0)=\{s_\mu(0)\}$. Generically, this vector can be 
written in the basis of eigenvectors of the $\mathbf J$ matrix as
\begin{equation}
\vec S(0) = \sum_\nu c_\nu \vec V_\nu
\; ,
\end{equation}
with the normalization  translating into the following condition on the coefficients $c_\nu$:
\begin{equation}
S^2(0) = \sum_{\nu\eta} c_\nu c_\eta \, \vec V_\nu \cdot \vec V_\eta = \sum_{\nu} c^2_\nu = N
\; . 
\end{equation}
Interesting choices for the initial state are: 

\begin{enumerate}
\item[(i)] A flat distribution on the basis of eigenvectors, that is $c_\nu=1$ for all $\nu$, which can be associated to 
thermal equilibrium at a very high temperature. This choice was made in the main body 
of~\cite{CuDe95a,CuDe95b}, where the dynamics of the $N\to\infty$ 
system were considered, and in~\cite{FyPeSc15} where finite $N$ corrections were studied. 
It corresponds to  
\begin{eqnarray}
s_\mu(0) 
= s^{\rm flat}_{\mu}(0)=1 \quad\qquad \forall \mu \in [\![1,N]\!]
\; .
\end{eqnarray}
\item[(ii)]  A ``projected'' initial system $\vec{S}(0)$ is almost perfectly aligned with an eigenvector 
  $\vec{V}_\alpha$, with just a small orthogonal random perturbation $\vec\varepsilon$.
It can be implemented as follows:
\begin{eqnarray}
\vec{S}(0)=\vec{\varepsilon} +\sqrt{(N-\vec{\varepsilon}\cdot \vec{\varepsilon}\,)}\;
 \vec{V}_\alpha \quad\quad \text{with} \quad\quad \vec{\varepsilon} \cdot \vec{V}_\alpha=0 \; \text{,}\quad \quad \text{and}\quad {\vec{\varepsilon}}^{\;2}\ll N
\; .
\label{eq:initial-proj}
\end{eqnarray}
Therefore, 
\begin{eqnarray}
s_{\mu (\neq \alpha)}(0) = s^{\rm proj}_{\mu (\neq \alpha)}(0)= \vec \varepsilon \cdot \vec V_\mu = \varepsilon_\mu
\qquad\quad
\mbox{and}
\qquad\quad
s_\alpha(0) = s^{\rm proj}_\alpha(0)= \sqrt{N-\varepsilon^2}
\; . 
\end{eqnarray}
The normalisation of $\vec{S}(0)$ is ensured for all $\vec \varepsilon$.
and ${\vec{\varepsilon}}^{\,2} \ll N$ implies $\sum_\mu \varepsilon_\mu^2 \ll N$. 
The case $\vec\varepsilon = \vec 0$, with $N\to\infty$ was 
also considered in~\cite{CuDe95a}. Random fluctuations close to the projected initial 
states can then be generated by choosing random $\vec \varepsilon$ with the conditions above.

We select in this paper a $\vec\varepsilon$ which yields the same projection of the initial condition on all the eigenvectors
$\vec V_\mu$: 
\begin{eqnarray}
\vec{\varepsilon}= X \sum_{\nu(\neq \alpha)}\vec{V}_\nu 
\qquad
\implies 
\qquad 
\varepsilon_\mu =\vec{\varepsilon} \cdot \vec V_\mu = X \quad \forall \; \mu \neq \alpha
\quad
\mbox{and}
\quad
\varepsilon_\alpha=0
\; , 
\label{eq;varepsilonmu}
\end{eqnarray}
and 
\begin{eqnarray}
s^{\rm proj}_{\mu(\neq \alpha)}(0)
= X
\qquad
\mbox{and}
\qquad
s^{\rm proj}_{\alpha}(0)= \sqrt{N-X^2 (N-1)}\approx\sqrt{N}
\; . 
\end{eqnarray}
We then consider $ X$   
to be a  Gaussian random variable with zero mean $\langle X\rangle_{i.c.}=0$ and $\langle X^2\rangle_{i.c.} = \varepsilon_{typ}^2 \ll 1$, 
properties that ensure 
$\varepsilon^2 \ll N$ since
$\varepsilon^2 = X^2 \sum_{\mu (\neq \alpha)} 1 = X^2 (N-1) \sim X^2 N$. 
In the following we will always refer to this choice of initial conditions as the  ``projected initial conditions''.

\end{enumerate}

Another relevant choice for $\vec{\varepsilon}$ would be to consider each component $\varepsilon_\mu$ as an i.i.d. Gaussian random 
variable with zero mean and variance $\varepsilon^2_{typ}$. However, the  
calculations are harder in this case and it was not worth pursuing them here.

\subsubsection{The interaction matrix}
\label{subsec:edge}

The largest eigenvalue of a finite size $N\times N$ Gaussian Orthogonal Ensemble (GOE)  
matrix ${\mathbf J}$ scales as $\lambda_N = 2J +J\zeta N^{-2/3}$ with 
$\zeta$ governed by the ($\beta=1$) Tracy-Widom distribution~\cite{TrWi1,TrWi2}.

The finite size corrections of the eigenvalue distribution 
function $\rho(\lambda)$ are especially important at its border, close to $\lambda_N$. The derivation of the 
exact finite size corrections at the edge of $\rho(\lambda)$ still remains open. 
In Refs.~\cite{Me91,PeSc15} an asymptotic scaling function  for 
$\lambda_N-\lambda=\mathcal{O}(N^{-2/3})$ has been derived and it reads
\begin{eqnarray}
\rho_{edge}(\lambda)=\left\{
    \begin{array}{lll}
        a N^{1/3} (\lambda_N-\lambda) \; , & \quad N^{-2/3}(\lambda_N-\lambda)\rightarrow 0 \; , 
               \vspace{0.2cm}
        \\
        \dfrac{1}{\pi} \sqrt{\lambda_N-\lambda} \; , & \quad N^{-2/3}(\lambda_N-\lambda)\rightarrow +\infty \; .  \\
    \end{array}
\right.
\end{eqnarray}
The exact value of the coefficient $a$ is not known.

The level spacings $g_\mu$  are defined as $g_\mu = \lambda_\mu-\lambda_{\mu-1}$.
The probability distribution of the first gap, $g_N \equiv \lambda_N - \lambda_{N-1}$, in the GOE was studied in~\cite{PeSc14,PeSc15} and it scales as
\begin{equation}
  \rho_{\rm gap}(g_N,N) = 
  N^{2/3} \ \rho_{\rm typ}(N^{2/3}g_N)
  \label{eq:first-gap1}
\end{equation}
where
\begin{equation}
 \left\{ 
    \begin{array}{rcl}
    \rho_{\rm typ}(N^{2/3} g_N) &\sim& b\,N^{2/3}g_N + o(N^{2/3}g_N)
       \vspace{0.3cm}
    \\
    \vspace{0.2cm}
    \ln{\rho_{\rm typ}(N^{2/3} g_N)} &\sim& -\frac{2}{3}   (N^{2/3} \,g_N)^{3/2} + o(N\,g_N^{3/2})
    \end{array}
    \right.
    \qquad
    \mbox{for}
    \qquad
    \begin{array}{l}
     N^{2/3}g_N \rightarrow 0
     \; , 
        \vspace{0.3cm}
     \\
        \vspace{0.2cm}
     N^{2/3}g_N \rightarrow \infty
     \; , 
    \end{array}
    \label{eq:gapdist}
\end{equation}
with $b$ another unknown constant. 

\begin{figure}[h!]
\begin{center}
\includegraphics[scale=0.3]{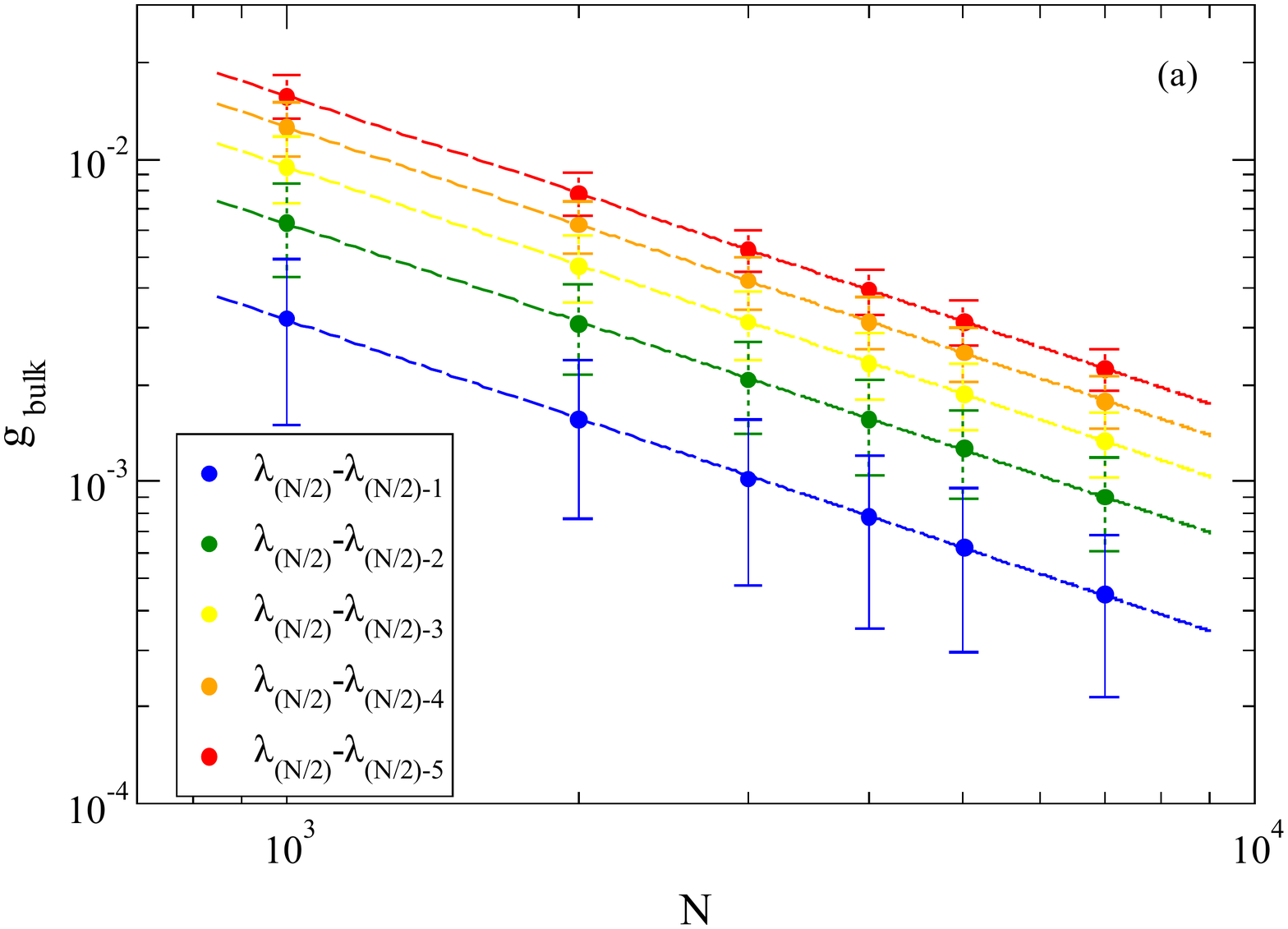}
\includegraphics[scale=0.3]{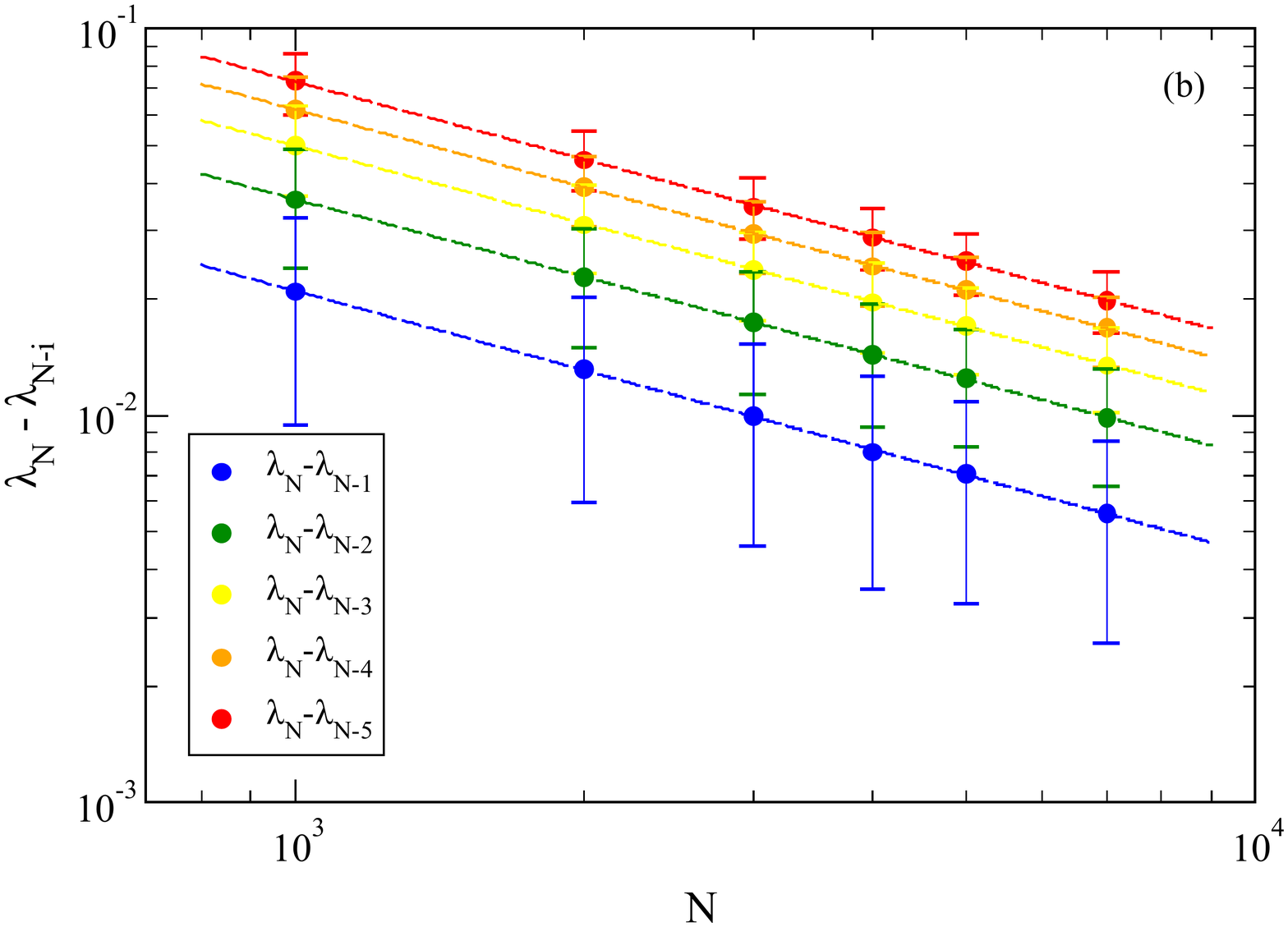}
\end{center}
\vspace{-0.35cm}
\caption{\small 
(a) The $N$ dependence of the disorder averaged level spacing in the bulk. Check of the 
expected scaling $\lambda_i - \lambda_{i-1} \simeq N^{-1}$, on average and evaluation 
of the standard deviation.
(b) The $N$ dependence of the disorder averaged distance to the largest eigenvalue, 
$d_i = \lambda_N - \lambda_{N-i}$
  for $i=1\ldots 5$. The first gap, $g_N$, corresponds to $i=1$.
  Averages were taken over $1024$ sets of eigenvalues of the $\mathbf J$ matrix for each $N$.
Details of the fits shown with dashed lines are shown in Table~\ref{tab:avgapstab}.}
\label{fig:avgaps}
\end{figure}

In Fig.~\ref{fig:avgaps}~(a) we check  a well-known result, the fact that $g_\mu=\lambda_\mu-\lambda_{\mu-1}$
in the bulk of the spectrum scales as $1/N$~\cite{FyLeDou14}. In the same panel we display some 
other distances $\lambda_\mu-\lambda_{\mu-\nu}$ with $\nu=2, \dots, 5$ and we verify that they all 
scale with $N$ in the same way. The data are averaged over ${\cal N} = 1024$ samples of eigenvalues
and the figure also shows the standard deviation.
The $N$ dependence of the disorder averaged first gap as well as the one of other distances to the 
largest eigenvalue, $d_i= \lambda_N-\lambda_i$,  are shown in
Fig.~\ref{fig:avgaps}~(b). 
Averages are also computed using ${\cal N} = 1024$ samples of eigenvalues.
All $d_i$ demonstrate an algebraic decay with  $N$. In Table \ref{tab:avgapstab} we see that the fit
exponent of the first gap, $g_N=d_1$, is slightly larger than the theoretically expected $2/3$, but also that the exponents seem to
approach continuously the value $2/3$ for higher order distances. 

\begin{figure}[h!]
\vspace{0.25 cm}
\begin{center}
\includegraphics[scale=0.65]{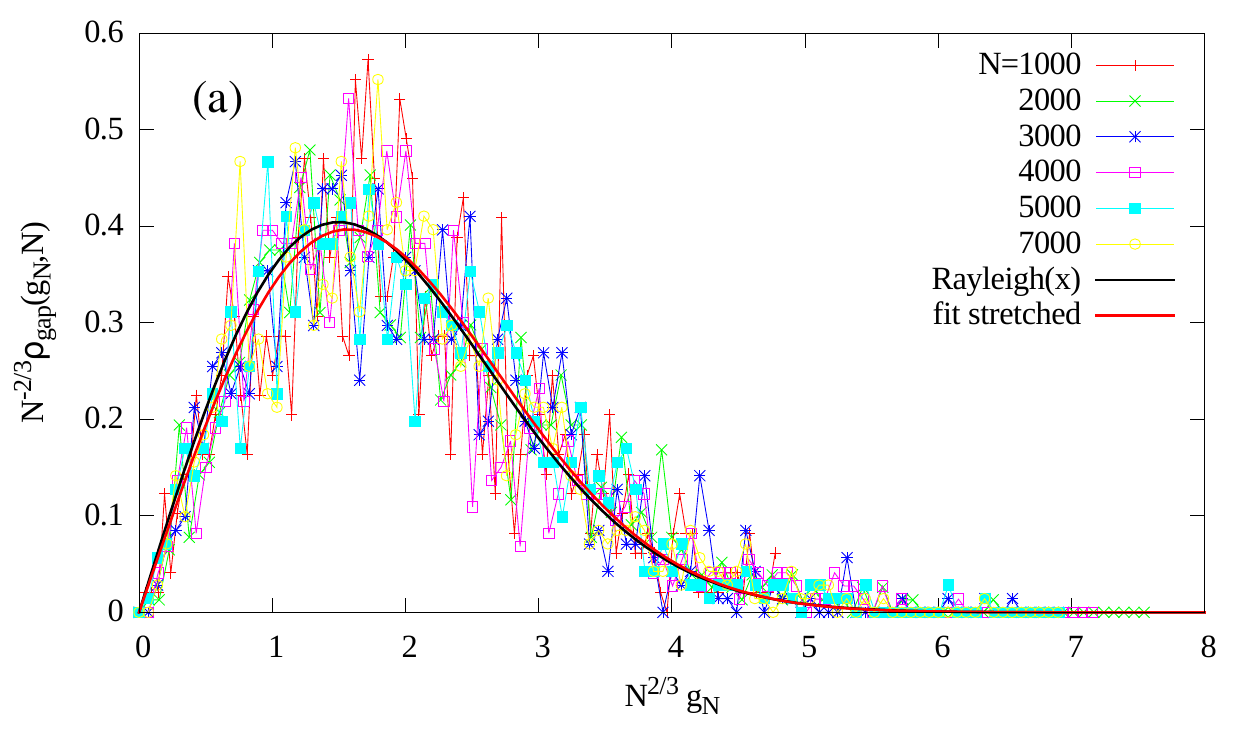}
\includegraphics[scale=0.65]{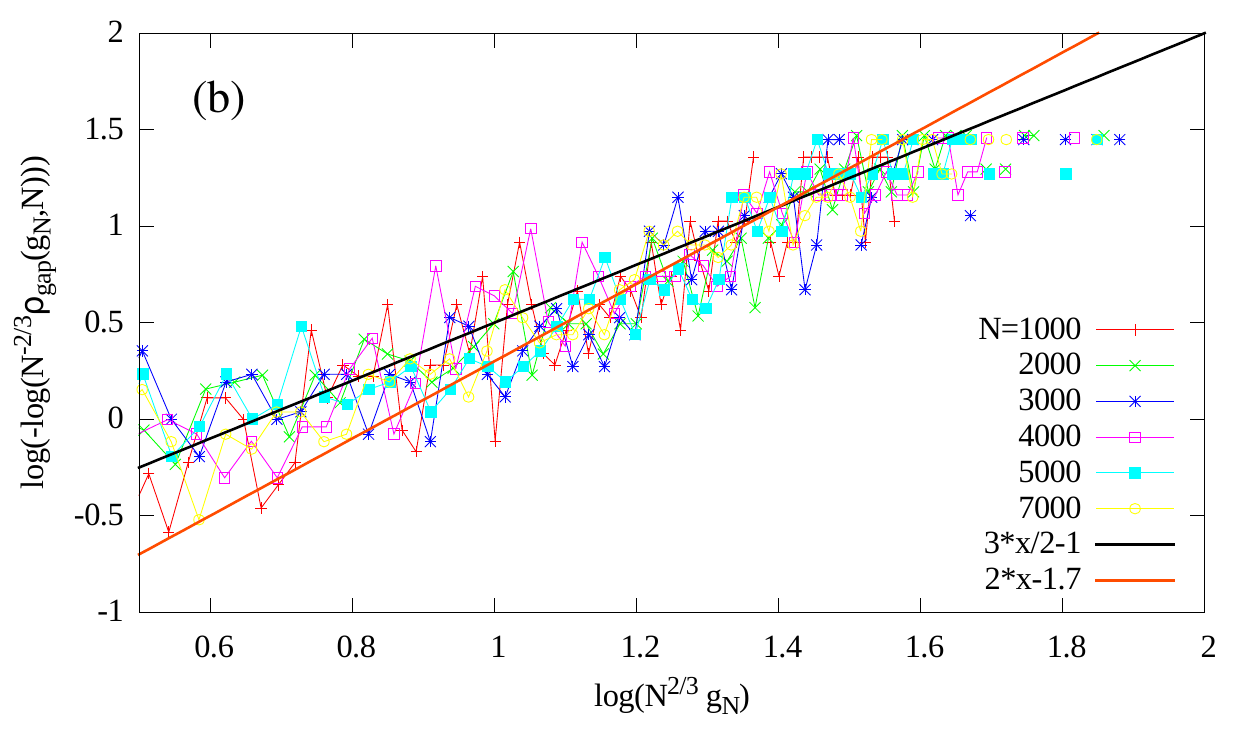}
\end{center}
\vspace{-0.35cm}
\caption{\small 
(a) Data collapse of the
  distribution of the first gap, $g_N$, for different system sizes according to eq.~(\ref{eq:first-gap1}).
  In black a Rayleigh distribution and in red a fit to the $N=7000$ data (see text for details).
  (b) The distribution of the extreme $g_N$ values, with the Rayleigh Gaussian tail (in red) and the 
  tail according to the Tracy-Widom behaviour~\cite{PeSc14} (in black).  }
\label{fig:pdf-gn}
\end{figure}

In the middle of the spectrum, the Wigner surmise predicts that the distribution of 
level spacings $g_\mu$ in the GOE ensemble takes the form of a Rayleigh distribution, 
with a linear increase from zero representing the level repulsion and a cross-over 
towards a Gaussian decay at large values of $g_\mu$:
\begin{equation}
  \rho_{R}(x)=\frac{x}{\sigma^2}\,e^{-x^2/(2\sigma^2)}
\end{equation}
In Fig.~\ref{fig:pdf-gn}~(a) we display 
the scaled probability distribution function of $g_N$ drawn from the same ensemble 
of ${\cal N}$ matrices, for the system sizes given in the key. The data are noisy 
and more samples would be needed to smooth it. The solid 
black curve is a Rayleigh distribution while the red  one is a fit to the $N=7000$ data
with a function of the form
$f(x)=a\,x^b\exp{(-c\,x^d)}$. The fit 
  yields values of the exponents $b$ and $d$ which are very close to the Rayleigh ones.
  In panel (b) we 
zoom over the tail of the distribution and we present, in red, the line that 
corresponds to the fit with the exponent $d=2$ representing the Gaussian tail, and 
in black the stretched exponential tail with exponent $d=3/2$ predicted in~\cite{PeSc14,PeSc15}.
Within the uncertainty of the raw data, the figure shows that the 
latter gives a better representation of  the data.
Having said so, it is known that deriving the 
full distribution of $g_N$ is a very difficult task. For the 
GUE ensemble, this was achieved in~\cite{WiBoFo13} 
and a more manageable expression allowing for an asymptotic analysis was derived in~\cite{PeSc14,PeSc15}.
The techniques developed in~\cite{MaPoSc21} may allow one to obtain the gap distribution analytically in the 
GOE as well.

\begin{table}[]
\centering
\begin{tabular}{cccc}
\hline
Gap                & \multicolumn{2}{c}{$y=a \times x^b$} & Standard Error \\ \hline
\multirow{2}{*}{$\lambda_N - \lambda_{N-1}$} & $a$                 & 2,3315                   & 0,03862183  \\
                   & $b$                 & -0,6817                  & 0,004794667 \\ \hline
\multirow{2}{*}{$\lambda_N - \lambda_{N-2}$} & $a$                 & 3,6211                   & 0,02553595  \\
                   & $b $                & -0,66656                 & 0,003170135 \\ \hline
\multirow{2}{*}{$\lambda_N - \lambda_{N-3}$} & $a  $               & 5,0813                   & 0,02722184  \\
                   & $b$                 & -0,66936                 & 0,003379427 \\ \hline
\multirow{2}{*}{$\lambda_N - \lambda_{N-4}$} &$ a  $               & 6,3192                   & 0,02236387  \\
                   & $b $                & -0,66971                 & 0,002776339  \\ \hline
\multirow{2}{*}{$\lambda_N - \lambda_{N-5}$} & $a $                & 7,3988                   & 0,02250885  \\
                   & $b  $               & -0,66851                 & 0,002794338 \\ \hline
\end{tabular}
\caption{Results of power law fits to the average distances from the largest eigenvalue shown in
  Fig.~\ref{fig:avgaps}(b).}
\label{tab:avgapstab}
\end{table}

We are also interested in finding the minimum of the first gap, $g_N$, 
within the set of such gaps drawn from different random matrices
 labeled by $k=1, \dots, {\cal N}$:
\begin{equation}
g_N^{\rm min}({\cal N}) = \min\limits_{k} g_N^{(k)}
\; .
\end{equation}
 In the strict  ${\cal N}\to \infty$ limit,
\begin{equation}
\lim_{{\cal N}\to \infty} g_N^{\rm min}({\cal N}) = 0
\; ,
\end{equation}
given that $\rho_{\rm gap}$ has support on $g_N = [0,\infty)$.
However, in typical samplings ${\cal N}$ is large but does not diverge and, therefore, $g_N^{\rm min}({\cal N})$ depends on 
${\cal N}$. Indeed, it decreases with 
${\cal N}$ in a way that we can estimate by assuming that there is typically one (or a few) value(s) $g_N^{(k)}$ in the interval 
$[0, \mu_{\cal N}]$, implying that \cite{Me91}
\begin{eqnarray}
 \frac{1}{{\cal N}}
&=&
\int_0^{\mu_{\cal N}} dg_N \; \rho_{\rm gap}(g_N)  
= 
\int_0^{\mu_{\cal N}} dg_N \; N^{2/3} \rho_{\rm typ}(N^{2/3} g_N)  = 
\int_0^{N^{2/3} \mu_{\cal N}} dy \; \rho_{\rm typ}(y) 
\nonumber\\
&\sim&
\int_0^{N^{2/3} \mu_{\cal N}} dy \;  b \,  y = 
\frac{b}{2} \; (N^{2/3} \mu_{\cal N})^2
\qquad
\implies 
\qquad
\mu_{\cal N} \sim N^{-2/3} \; {\cal N}^{-1/2}
\; . 
\end{eqnarray}
In this way, we signalled out  from the rest the minimal value we are looking for; it scales as 
\begin{equation}
g_N^{\rm min}({\cal N}) \sim N^{-2/3} \; {\cal N}^{-1/2}
\; . 
\label{eq:scaling-gNmin}
\end{equation}

\begin{figure}[h!]
\begin{center}
\includegraphics[scale=0.3]{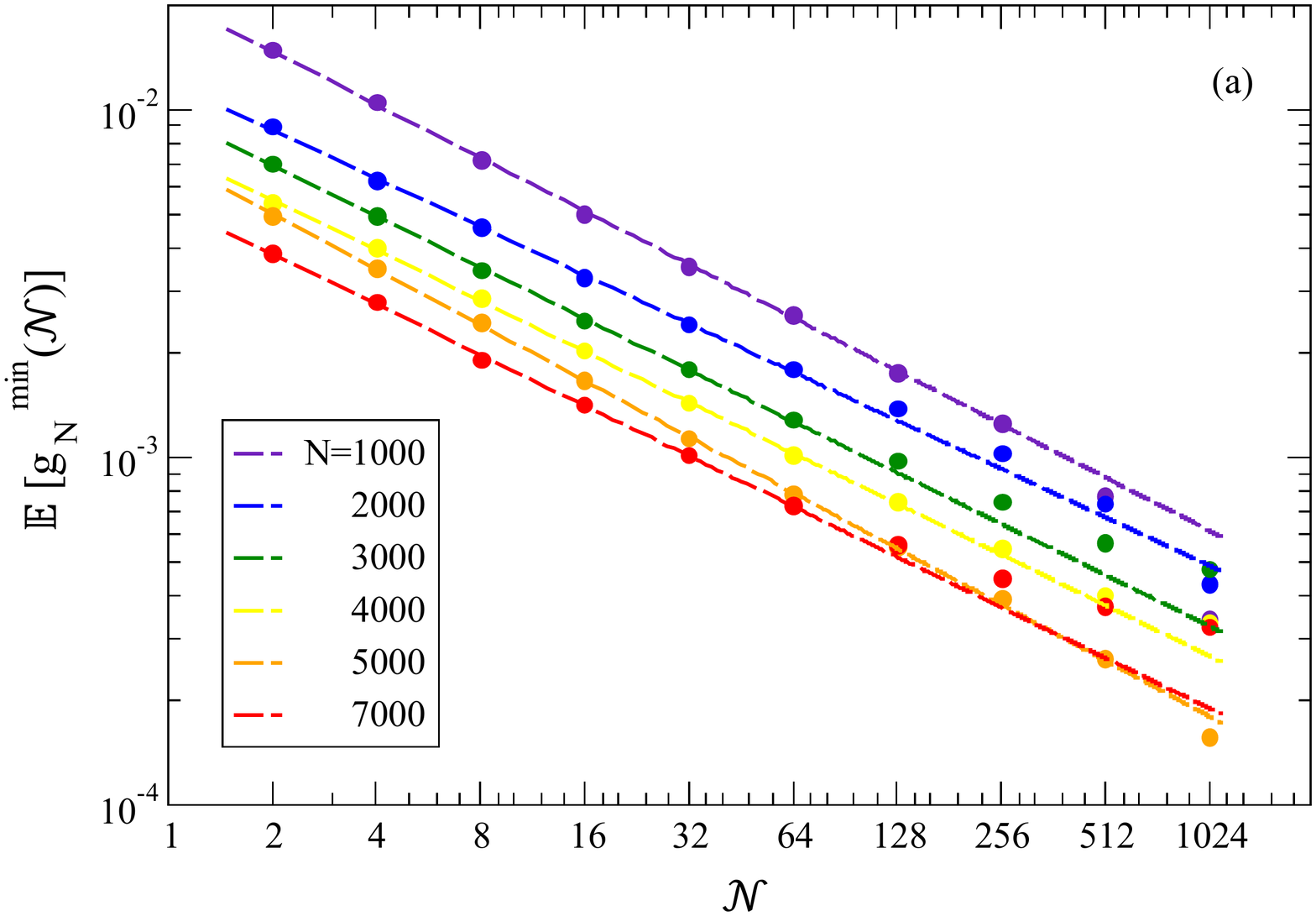}
\includegraphics[scale=0.3]{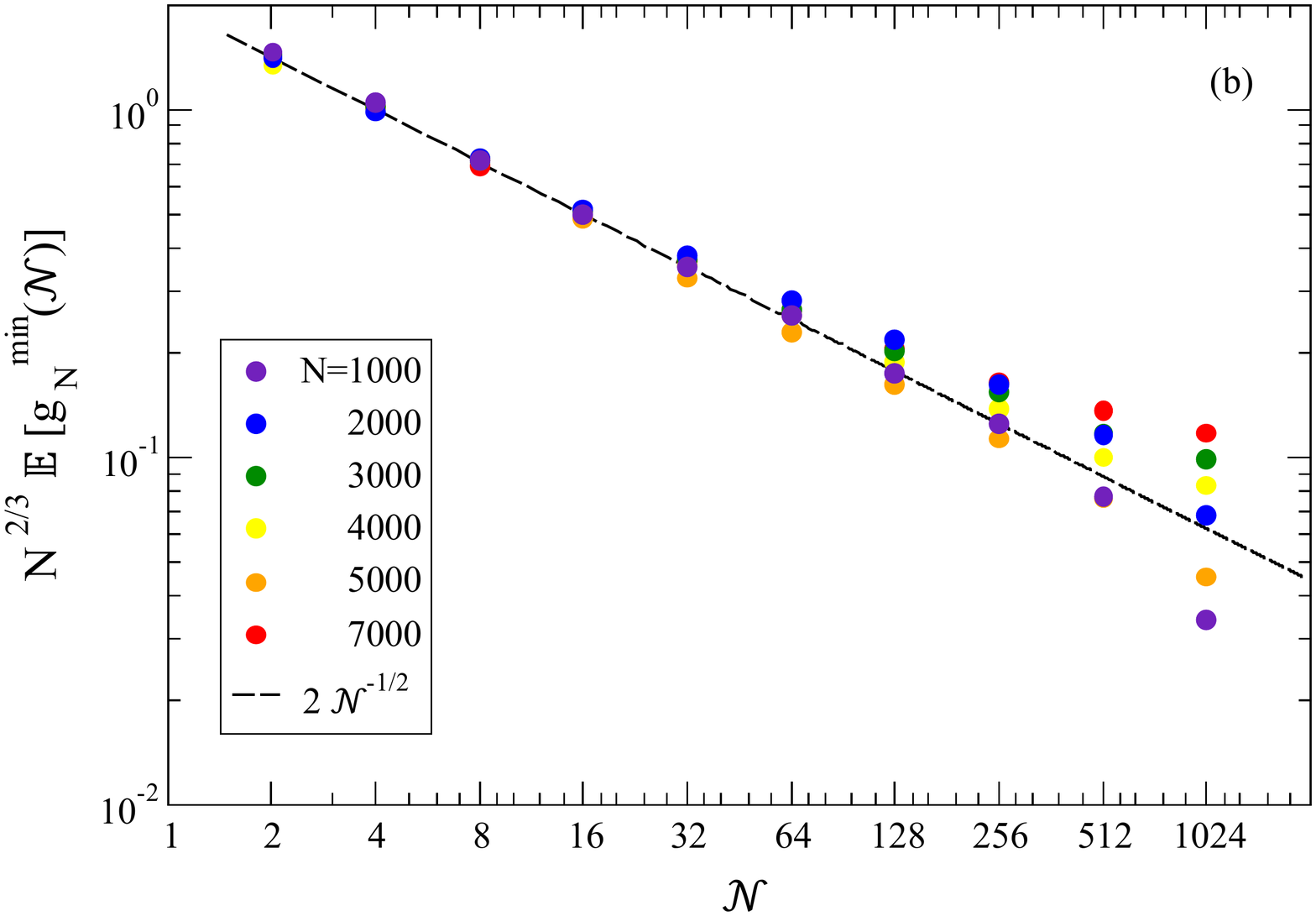}
\end{center}
\caption{\small (a) The ${\mathcal N}$ dependence of the average minimal gap for random matrix sizes $N=1000-7000$
  as indicated in the key. For each ${\mathcal N}$ an average over $1000$ random but correlated samples sorted from
  $1024$ uncorrelated ones was done.
  The dotted lines are fits to the data in the range $[2,64]$, where correlations
  are still not severe .
  (b) The same data as in panel (a) but showing the collapse expected according to equation (\ref{eq:scaling-gNmin})
  (see text for details). The dashed line is an interpolation with the expected power law.}
\label{fig:mingaps}
\end{figure}

\begin{table}[h!]
\centering
\begin{tabular}{cccc}
\hline
$N$                & \multicolumn{2}{c}{$y=a \times x^b$} & Standard Error \\ \hline
\multirow{2}{*}{$1000$} & $a $                & 0.020956                   & 0.006281549  \\
                   & $b $                & -0.50896                 & 0.01695654 \\ \hline
\multirow{2}{*}{$2000$} & $a  $               & 0.012077                   & 0.005501834  \\
                   & $b$                 & -0.46281                 & 0,01485176 \\ \hline
\multirow{2}{*}{$3000$} & $a  $               & 0.0097188                 & 0.005044484   \\
                   & $b $                & -0.48937                 & 0.01361718 \\ \hline
\multirow{2}{*}{$4000$} & $a   $              & 0.0077346                   & 0.006292685   \\
                   & $b  $               & -0.48489                 & 0.0169866  \\ \hline
\multirow{2}{*}{$5000$} & $a      $           & 0.0073124                   & 0.005300066  \\
                   &$ b    $             & -0.53454                 & 0.01430711 \\ \hline
\multirow{2}{*}{$7000$} & $a  $               & 0.0053732                   & 0.007193261  \\
                   & $b   $              & -0.48237                 & 0.01941763 \\ \hline                   
\end{tabular}
\caption{Results of power law fits to the average minimal gaps shown in panel (a) of Fig. \ref{fig:mingaps}.}
\label{tab:mingapstab}
\end{table}

In Fig.~\ref{fig:mingaps} we show numerical results. A problem to obtain reliable numerical data on
the minimal gaps is that a huge set of random matrices is needed. Note that a single $g_N^{\rm min}$ refers to the minimal
value among a set of ${\cal N}$ random matrices, from each of which we extract the first gap $g_N$. In order to
compute average values and fluctuations a large set of matrices needs to be diagonalized. For large matrix sizes
$N$ this is a computationally demanding task. We decided to compute the average $g_N^{\rm min}$ for different matrix
sizes by a resampling procedure. From an original set of $1024$ independent matrices we computed averages from $1000$
random sets of ${\cal N}$ matrices each. Of course, these sets are correlated.
In panel (a) we see the results of the resampling procedure. For each system size $N$ the average $g_N^{\rm min}$ is
shown together with a power law fit including the data for ${\cal N}\leq 64$. In Table \ref{tab:mingapstab} we show
the results of the fits for each size. The fit exponents are not far from the expected value $1/2$. Furthermore,
the prefactors should show a scaling with system size $N^{-2/3}$.
In panel (b) of Fig.~\ref{fig:mingaps} we verify that sets
with ${\cal N}\leq 64$ show a good data collapse 
in agreement with eq. (\ref{eq:scaling-gNmin}). For larger set sizes deviations from the expected decay grow
fast due to the strong correlations between sets for our limited number of independent samples.

\subsection{The energy and the Lagrange multiplier}

The fluctuating energy density,
\begin{equation}
e(t,N) = -\frac{1}{2N} \sum_{i\neq j} J_{ij} s_i(t) s_j(t) = -\frac{1}{2N} \sum_\mu \lambda_\mu s_\mu(t) s_\mu(t)
\; , 
\end{equation}
is simply related to the fluctuating Lagrange multiplier $z$. This can be proven as follows. Multiplying the Langevin 
equation by $s_\mu(t')$, summing over $\mu$, and taking $t'\to t$ one has
\begin{equation}
  \frac{1}{N}  \sum_\mu (d_t s_\mu(t)) s_\mu(t' \to t) = 2e(t,N) + z(t,N)
  \; . 
  \label{eq:asymptotic-z-energy}
\end{equation}
In the left-hand-side we identify the time-derivative of the self-correlation 
\begin{equation}
   C(t,t')=\frac{1}{N}\sum_{i} s_i(t)s_i(t') = \frac{1}{N}\sum_{\mu} s_\mu(t)s_\mu(t')
   \label{eq:autocorrel}
 \end{equation}
 evaluated at equal times. Since it should vanish for all $N$  as a consequence of the 
normalisation $C(t,t) = 1$, one then has
\begin{equation}
  z(t,N)=- 2e(t,N)
  \label{eq:zener}
\end{equation}
for any  realisation of the random matrix and for all initial conditions.


An exact expression for $z$, and $e$,  can next be deduced.
The over-damped dynamic equation~(\ref{eq:langevin-mu}) is deterministic and its solution fixed by the initial condition.
After a straightforward integration, the time-dependent spin configuration is given by 
\begin{eqnarray}
s_\mu(t)=s_\mu(0)\exp{\Big[\lambda_\mu t-\int_0^t dt'\,z\big(t',\{s_\mu(0)\}\big)\Big]}
\, ,
\label{eq:smu-t}
\end{eqnarray}
where we stressed that the Lagrange multiplier depends on the initial condition. In order
to fix it, we impose  the spherical constraint
\begin{eqnarray}
&& N=\sum_\mu s^2_\mu(t)=\exp{\Big[-2\int_0^t dt' \, z\big(t',\{s_\mu(0)\}\big)\Big]}\sum_\mu s^2_\mu (0)\; e^{2\lambda_\mu t}
\nonumber\\
&& \qquad\qquad 
\implies 
\quad 
2  \int_0^t dt'\, z\big(t',\{s_\mu(0)\}\big)=\ln\Big[\frac{1}{N}\sum_\mu s^2_\mu (0) \;e^{2\lambda_\mu t}\Big]
\nonumber\\
& &
\qquad\qquad 
\implies 
\quad 
z\big(t,\{s_\mu(0)\}\big)=\frac{\sum_\mu  s^2_\mu(0) \; \lambda_\mu \; e^{2\lambda_\mu t}}{\sum_\mu s^2_\mu(0)\; e^{2\lambda_\mu t}}
=
\frac{1}{2} \frac{d}{dt} \ln\Big[ \frac{1}{N} \sum_\mu  s^2_\mu(0)  \; e^{2\lambda_\mu t}\Big]
\label{eq:fixing-z}
\, .
\end{eqnarray}
Separating the contribution from the largest eigenvalue, 
\begin{equation}
z\big(t,\{s_\mu(0)\}\big)=
\lambda_N + \frac{1}{2}  \frac{d}{dt} \ln {\mathcal Z}_N\big(t,\{s_\mu(0)\}\big)
\label{eq:fixing-z-bis}
\end{equation}
with 
\begin{equation}
{\mathcal Z}_N\big(t,\{s_\mu(0)\}\big)
=  \frac{s^2_N(0)}{N} + \frac{1}{N} \sum_{\mu(\neq N)}  s^2_\mu(0) \; e^{2(\lambda_\mu-\lambda_N) t}
\end{equation}
and ${\mathcal Z}_N\big(0,\{s_\mu(0)\}\big)=1$ for all initial conditions.

For finite size systems, any initial state that is not fully aligned with an eigenvector of the interaction matrix, 
and regardless of the scaling of time $t$ with $N$, the Lagrange multiplier converges to $\lambda_N$ when $t\rightarrow \infty$. 
Indeed, in the previous equation the two sums will eventually be dominated by the contribution of the largest eigenmode 
as long as the initial condition has some overlap with it. As an example
\begin{eqnarray}
\sum_\mu s^2_\mu(0)\;e^{2\lambda_\mu t}\underset{t\rightarrow \infty}{=}s^2_{N}\;e^{2\lambda_N t}
\end{eqnarray}
and it follows that
\begin{eqnarray}
z\big(t,\{s_\mu(0)\}\big)&\underset{t\rightarrow \infty}{=}&\Big[\lambda_N s_N^2(0) \;e^{2\lambda_N  t}
\Big]\Big[s^2_N(0)\;e^{2\lambda_N  t}\Big]^{-1}=\lambda_N \; , \qquad\quad \forall \; \mbox{finite} \; N 
\label{eq:zlambdamax}
\end{eqnarray}
and for all initial conditions with $s^2_N(0) \neq 0$.

Going back to eq.~(\ref{eq:fixing-z-bis}), this implies that the excess energy with respect to the ground state one is
\begin{eqnarray}
\Delta e(t,N) 
&=&\lambda_N-e(t,N)\nonumber\\
&=&\frac{1}{2} \; \frac{\sum\limits_{\mu(\neq N)}  \; s_\mu^2(0) (\lambda_N-\lambda_\mu) \; e^{2(\lambda_\mu-\lambda_N) t}}{s_N^2(0)+\sum\limits_{\mu(\neq N)}  s_\mu^2(0) e^{2(\lambda_\mu-\lambda_N) t}}
\; . 
\end{eqnarray}
In the very last time-regime,  assuming that only one term dominates the sums:
\begin{equation}
\Delta e(t,N) \to \frac{g_N s_{N-1}^2(0)}{2s_{N}^2(0)} \; e^{-2 g_N t}
\; ,
\label{eq:last-exp}
\end{equation}
with the gap between the last two eigenvalues defined as $g_N = \lambda_N - \lambda_{N-1}$. 
(The trivial case in which the initial configuration is correlated with the ground state and  $s_N^2(0)$ scales with $N$ is excluded.)

The projection of the time-dependent spin vector $\vec S(t)$ on any of the eigenvectors $\vec V_\mu$, $s_\mu(t) = \vec S(t) \cdot \vec V_\mu$, 
reads
\begin{equation}
s_\mu(t) = 
 s_\mu(0)   \; \exp\left[ \lambda_\mu t - \int_0^t dt' \; z(t', \{s_\mu(0)\}) \right]
= 
s_\mu(0)   \; \exp\left[ (\lambda_\mu -\lambda_N) t + 2\int_0^t dt' \;\Delta e (t', \{s_\mu(0)\}) \right]
\; . 
\end{equation}

\section{Infinite temperature initial states}
\label{sec:flat}

In this section  we study the fluctuations with respect to the random matrix  ${\mathbf J}$. 
The initial conditions are drawn from the flat distribution (i) defined in Sec.~\ref{subsec:initial}.
The disorder averaged asymptotic ground state energy density of the infinite size system is expected to be equal to the one 
of any random couplings realization (self-averageness) and approach~\cite{CuDe95a}
\begin{equation}
\lim_{t\gg 1} \lim_{N\to\infty} {\rm I\!E} [e(t,N) ]= -1 + \frac{3}{8 t}
\; ,
\label{eq:asympt-energy}
\end{equation}
where the first term in the r.h.s. is $- \lim_{N\to\infty} \lambda_N/2$ and 
henceforth we measure the energy in units of $J$.
 
For any $N$ and for fixed random interactions, from eqs.~(\ref{eq:zener}) and (\ref{eq:zlambdamax}), 
$\lim_{t \to \infty}e(t,N) =-\lambda_N/2$. For $N$ large but finite Fyodorov {\it et al.} proved the existence of 
a cross-over time-scale $t_{\rm cross} = {\mathcal O}(N^{2/3})$
between two regimes~\cite{FyPeSc15}. For $t\ll t_{\rm cross}$ the dynamics is well described by the one of the infinite size system. Instead,
beyond this time scale, the system explores a sector of the potential energy landscape close to the ground state, and the 
relaxation is governed by a small number of saddle points with a number of unstable directions of $ {\mathcal O}(1)$.

\begin{figure}[h!]
\begin{center}
\includegraphics[scale=0.40]{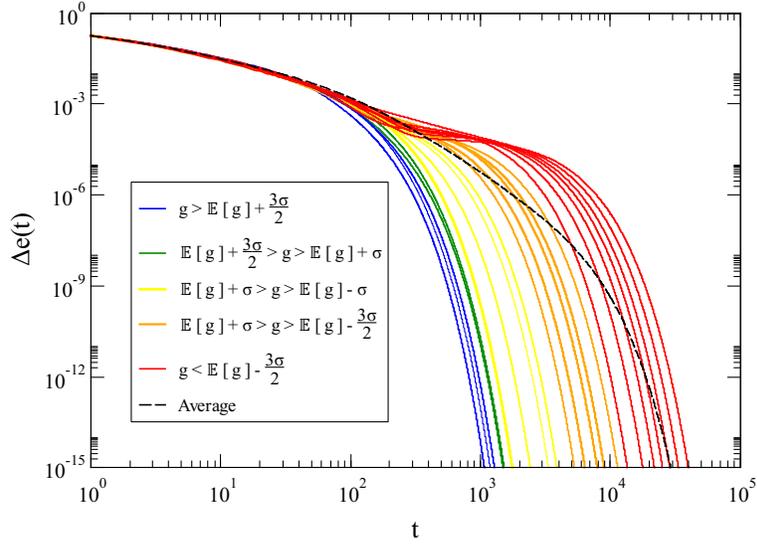}
\end{center}
\vspace{-0.2cm}
\caption{\small Random matrix induced fluctuations of the excess energy in systems with $N=7000$ evolving from
  random initial conditions. Decay of the excess energy for systems with different realisations of the coupling
  strengths and, with a dashed black line, the average over 1024 such runs. Note the grouping of 
  curves with different colour according to the deviation of the first gap $g_N$ from its average value
  (see the text for discussion).}
\label{fig:energies7000}
\end{figure}

More precisely, in the finite $N$ system
the mean excess energy density with respect to the ground state,
${\rm I\!E} [\Delta e(t,N) ] ={\rm I\!E} [e(t,N)+\lambda_N /2]$, behaves as
\begin{equation}
    {\rm I\!E} [\Delta e(t,N) ] \sim \left\{ 
    \begin{array}{l}
    e_1(t) 
    \\
    N^{-2/3} \; f\left(t N^{-2/3}\right) 
    \end{array}
    \right.
    \qquad
    \mbox{for}
    \qquad
    \begin{array}{l}
     t\ll N^{2/3} 
     \\
t\geqslant N^{2/3}
    \end{array}
    \label{eq:gapscaling1}
\end{equation}
where the function $e_1(t)=3/(8t)$ makes contact with~eq.~(\ref{eq:asympt-energy}),
and $f(x)$ is a scaling function with the limits:
\begin{equation}
  f(x) \sim 
  \left\{ \begin{array}{l}
    \dfrac{3}{8x} 
    \vspace{0.2cm}
    \\
    \dfrac{a}{x^3} 
  \end{array}
  \right.
  \qquad \mbox{if} \qquad
\begin{array}{l}
 x \to 0 
 \vspace{0.5cm}
 \\
 x \to \infty
  \end{array}
    \label{eq:gapscaling2}  
\end{equation}
$a$ is an unknown constant. The regime 
$x\to 0$ corresponds to the thermodynamic limit, while $x\to\infty$ describes the late time behaviour 
of systems with large  but finite $N$. Therefore, for 
$t > t_{\rm cross}$, the system escapes the self-averaging algebraic decay and enters another,  faster and non self-averaging, but 
also algebraic decay. These features can be seen in Fig.~\ref{fig:energies7000}. 

Figure~\ref{fig:energies7000} demonstrates the non self-averageness of the relaxation dynamics, at time-scales beyond $N^{2/3}$,  
in quenches from fully disordered initial conditions.
The various curves represent the decay of the excess energy density in systems with size $N=7000$ for different realisations
of the random matrix $\mathbf J$. After a self-averaging period that coincides with the regime in which the behaviour
is the same as in the $N\to\infty$ limit, some instances start departing from the average, dominate it, and render 
the average algebraic though with a different power.
In the figure, the different instances are grouped with different colours depending on the deviation of the first gap,
$g_N$, from its average value. Note that the group where the gaps are the largest decay fast. Because of this fact these
samples do not contribute to the second averaged algebraic regime (nor to the third and final regime).
It can be further  noticed that, at sufficiently long times, the average relaxation becomes still faster. In fact, this
last decay towards the global minimum of the energy is exponential and ruled by the minimum gap, $g_N^{\rm min}$,
as will be shown below.

\begin{figure}[h!]
\vspace{-0.25cm}
\begin{center}
\includegraphics[scale=0.30]{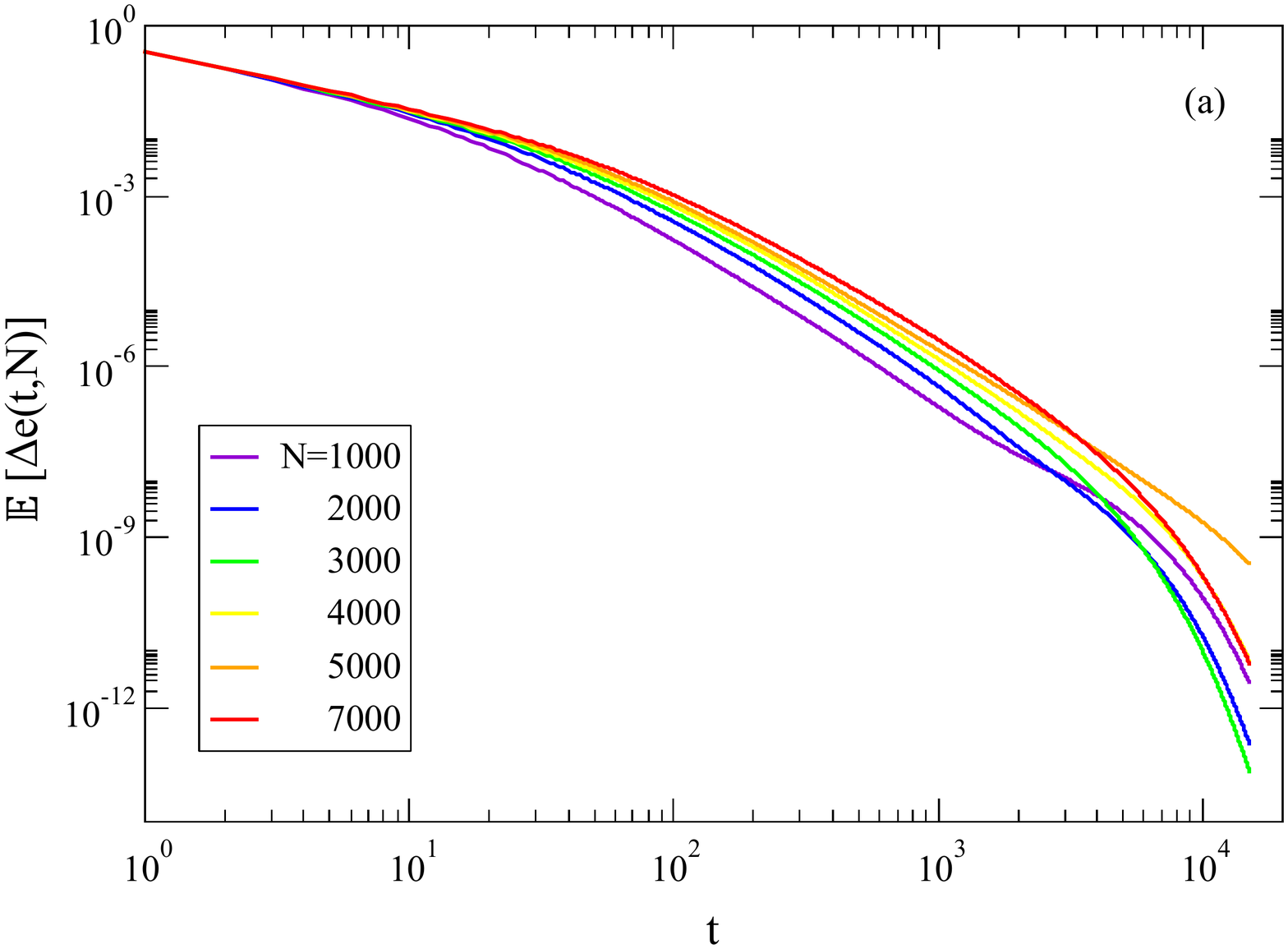}
\includegraphics[scale=0.30]{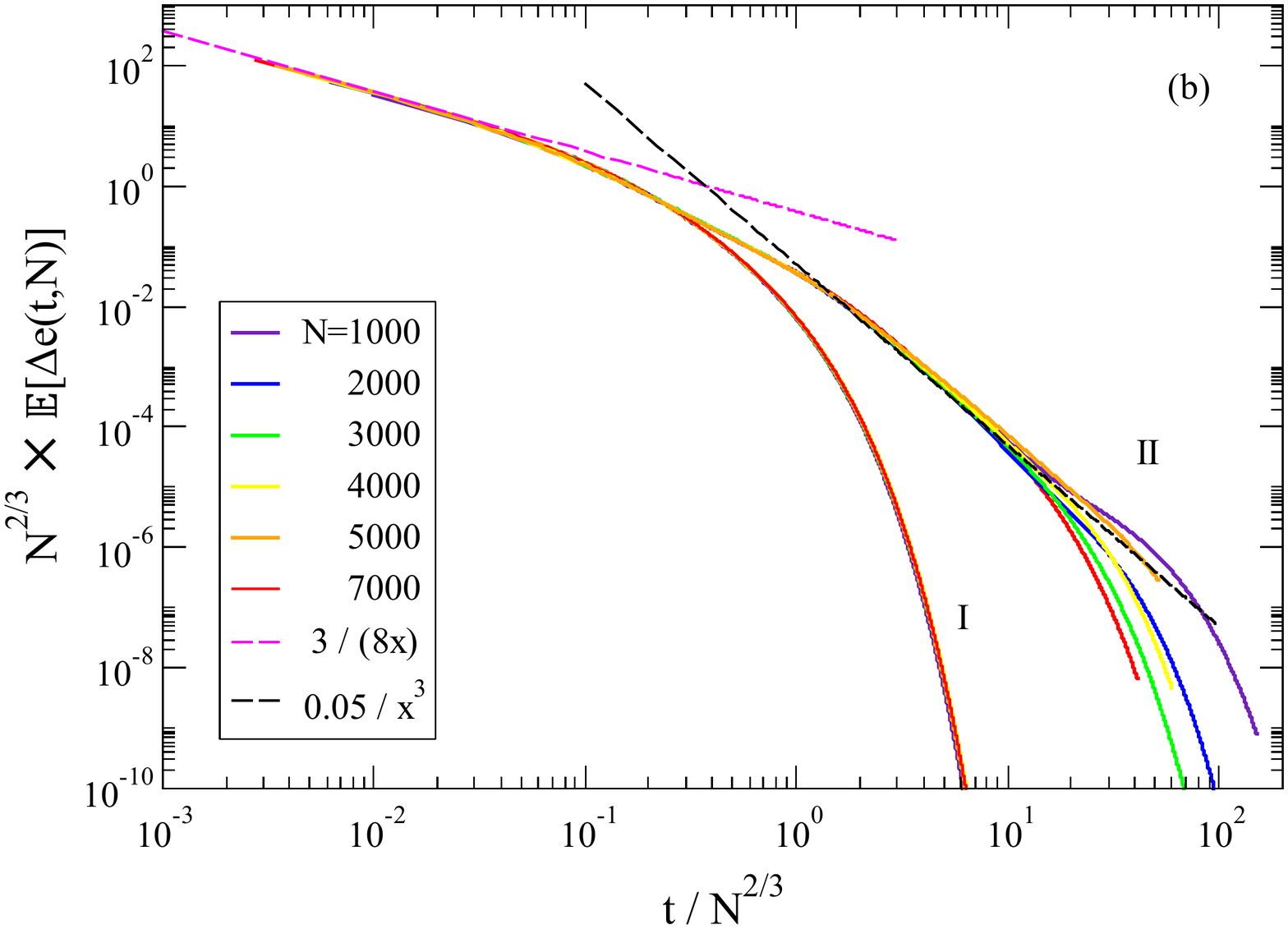}
\end{center}
\vspace{-0.25cm}
\caption{\small Finite-size dependence of the disorder averaged excess energy with respect to the ground state,
  $\Delta e$, in systems evolving from random initial conditions.
(a) Averages over 1024 disorder samples for different system sizes given in the key.
(b) Scaling according to eqs.~(\ref{eq:gapscaling1})-(\ref{eq:gapscaling2}). In set I samples with
$g_N > {\rm I\!E}(g_N)-\sigma$ and in set II the samples with smaller gaps.
The first two time regimes follow the analytic predictions in \cite{CuDe95a} and \cite{FyPeSc15}, respectively.
The spreading of the curves at the longest times in set II is an indication of the breakdown of this scaling. 
}
\label{fig:energiessizes}
\end{figure}

Figure~\ref{fig:energiessizes} shows the decay of the disorder averaged excess energy for different system sizes. Each
curve corresponds to an average over $1024$ samples of the matrix $\mathbf J$. The curves on the left panel
are shown again in the right one after scaling according to eqs.~(\ref{eq:gapscaling1})-(\ref{eq:gapscaling2})~\cite{FyPeSc15}.
Similarly to what was done in Fig.~\ref{fig:energies7000}, the averages were taken after dividing individual
samples in two groups: in group I samples with $g_N > {\rm I\!E}(g_N)-\sigma$ and in group II samples
with smaller gaps.
Clearly, the set with the smaller gaps is responsible for the
second power law regime, while the larger gaps lead to fast, exponential decay. This set does not contribute
to the averages in the second time scale. Power law fits, shown in dotted straight
lines, are in good agreement with the theoretical predictions for the first two dynamical regimes, the one 
that coincides with the $N\to\infty$ results and the next algebraic one.
The latter regime is governed by the statistical properties of the first gap $g_N=\lambda_N-\lambda_{N-1}$. From
eq.~(\ref{eq:last-exp}) we obtain
\begin{equation}
  {\rm I\!E} [\Delta e(t,N)] \ \underset{t\rightarrow \infty}{\longrightarrow} \
  {\rm I\!E} \left[\frac{g_N}{2} \; e^{-2 g_N t}\right].
\end{equation}
Using now the probability distribution of the first gap~\cite{PeSc15} recalled in eqs.~(\ref{eq:first-gap1})-(\ref{eq:gapdist}), one 
can perform the average
\begin{eqnarray}
  {\rm I\!E} [\Delta e(t,N)] &=&  \int_0^\infty dr \; \frac{r}{2} \ e^{-2rt}\ \rho_{\rm gap}(r,N) 
  \nonumber\\
  &\sim& 
   \int_0^{N^{-2/3}} dr \; \frac{r}{2} e^{-2rt}\ b N^{4/3}  r +  \int_{N^{-2/3}}^\infty dr \; \frac{r}{2} e^{-2rt}\ N^{2/3} e^{-\frac{2}{3} (N^{2/3}  r)^{3/2}}
   \nonumber\ \\
  &\sim& 
 \frac{b}{2} N^{4/3} t^{-3} \int_0^{t/N^{2/3}} dx \; x^2 \ e^{-2x} 
  +\frac{ 1}{2 N^{2/3}}  \int_1^\infty dx \ x \ e^{-2t N^{-2/3} x} \ e^{- \frac{2}{3} x^{3/2}}
  \; . 
  \end{eqnarray}
In the regime $t/N^{2/3} \gg 1$ the first term dominates, the second one is negligible, and the result
can be re-arranged in the   scaling form:
  \begin{eqnarray}
   {\rm I\!E} [\Delta e(t,N)]
  &\sim& \frac{b}{8N^{2/3}}  \left(\frac{N^{2/3}}{t}\right)^3 
    \; ,
    \label{eq:scaling-2nd-regime}
\end{eqnarray}
consistently with the fact that the average is dominated by the averaged contribution of the smaller gaps, that is to say, the 
regime $N^{2/3} g_N \to 0$ in eqs.~(\ref{eq:gapdist}). This is the second algebraic regime studied in~\cite{FyPeSc15}.

Finally, we see a third and last time regime in which the averaged $ {\rm I\!E} [\Delta e(t,N)]$ is controlled by the
smallest gap found within the ensemble of random matrices used, $g_N^{\rm min}$. In this regime
\begin{equation}
 {\rm I\!E} [\Delta e(t,N)] \sim \frac{g_N^{\rm min}}{2} e^{-2 g_N^{\rm min} t}
 \label{eq:gmin}
 \; . 
 \end{equation}
Since we expect 
$g_N^{\rm min} \simeq {\cal N}^{-1/2} N^{-2/3}$, see eq.~(\ref{eq:scaling-gNmin}), the dependence on $N$ 
in the last regime is of the same form as the one in the second one, with $x=t/N^{2/3}$ the scaling variable.

\begin{figure}[h!]
\vspace{-0.25cm}
\begin{center}
\includegraphics[scale=0.30]{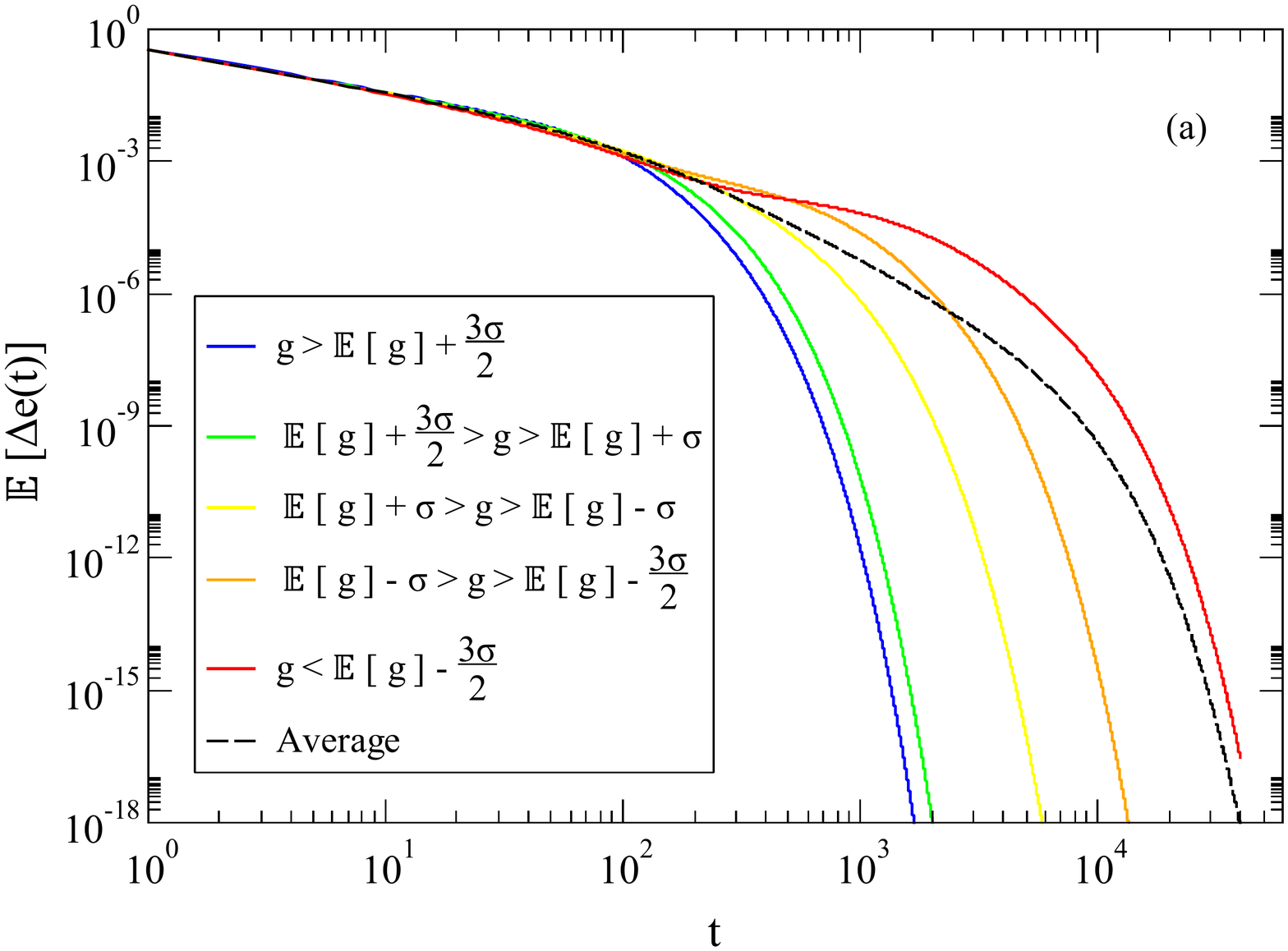}
\includegraphics[scale=0.30]{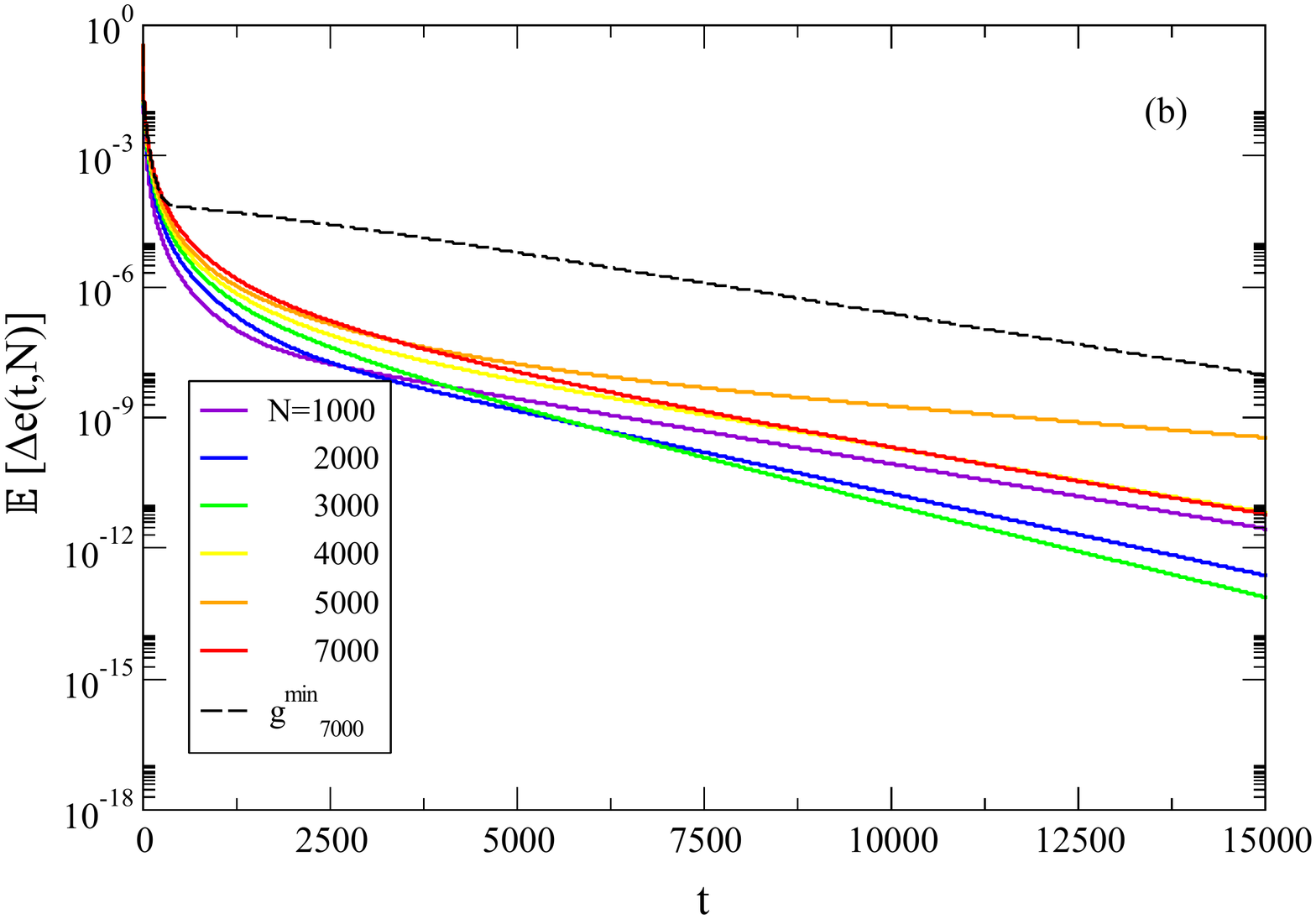}
\end{center}
\vspace{-0.25cm}
\caption{\small 
The averaged excess energy illustrating the behaviour of the longest time regime.
(a) Averages over subsets of samples grouped according to the distance of the first gap
relative to the average for $N=7000$.
(b) Averages for different sizes $N$ together with the decay of the sample with the minimal gap
for $N=7000$.
 }
\label{fig:avenerexponen}
\end{figure}

Figure~\ref{fig:avenerexponen}(a) illustrates this behaviour in the $N=7000$ case:
the curves with different colours are averages
over subsets of samples with the first gap in different ranges relative to the average gap. The curve on the
extreme right, in red, is an average on the set of samples with the smaller gaps, $g < {\rm I\!E}(g)-
3\sigma/2$, where $\sigma$ is the standard deviation of the first gap. The dotted black line corresponds
to the average excess energy over the whole set of $1024$ samples. It approximates asymptotically to the
rightmost curve, meaning that, as time grows, sets of samples with larger gaps gradually cease to contribute
to the average.

In Figure ~\ref{fig:avenerexponen}(b) the average excess energies for different system sizes are shown in
log-linear scale, together with the data for the sample corresponding to the smallest gap, $g_{7000}^{min}$,
for the $N=7000$ case (dotted black line). First of all, this scaling of the axes shows that at the longest times
the decay approaches an exponential behaviour.
The crossing of the curves for different system sizes reflects the large fluctuations
in the value of the smallest gaps for each $N$. Finally, we note that the dotted line
corresponding to the minimal gap for $N=7000$ goes, at the time scales shown, around three orders of
magnitude above the curve for the average excess energy. This shows that, although at asymptotic long
times there are a few exponentials contributing the the average excess energy, the approach to the longest
time scale given by $1/g_N^{min}$ is approached very slowly. This is expected, given that the differences
between the largest eigenvalue of the coupling matrix and the higher order neighbours scales similarly to
the first gap, $\lambda_N-\lambda_{N-i} \propto N^{-2/3}$, as shown in Fig.~\ref{fig:avgaps}(b) and
Table~\ref{tab:avgapstab}.

Finally, the crossover time between the second and the third regime should be determined by 
\begin{equation}
x^{-3}
\sim 
{\cal N}^{-1/2}  e^{-2 {\cal N}^{-1/2} x} 
\end{equation}
with $x=t/N^{2/3}$. The solution increases (slowly) with increasing ${\cal N}$ and diverges, 
as expected, in the ${\cal N}\to\infty$ limit.

Summarizing the results for disordered initial conditions, we have seen that finite sized systems show
three dynamical regimes. Two of them lead to algebraic relaxation. The first one is the trivial mean-field one, in which 
$N$ is so large that times can be considered finite with respect to it~\cite{CuDe95a}. The second one was 
identified  and described in~\cite{FyPeSc15}. We
have firmly confirmed the theoretical expectations with numerical results for large system sizes. The last
regime is exponential. We have shown that its scaling properties can be described exploiting the properties of the 
minimal gap drawn in the random matrix sampling, and hence depends on the size of the sampling matrix ensemble.

We can now examine 
the projection of the time-dependent spin vector on any of the eigenvectors, $s_\mu(t) = \vec S(t) \cdot \vec V_\mu$, 
\begin{equation}
s_\mu(t) = 
s_\mu(0)   \; \exp\left[ (\lambda_\mu -\lambda_N) t + 2\int_0^t dt' \;\Delta e (t', \{s_\mu(0)\}) \right]
\; . 
\end{equation}
In the first regime we can exploit that the fluctuations of $\Delta e$ with respect to its expected value 
are negligible and then write
\begin{equation}
s_\mu(t) \simeq
s_\mu(0)   \; \exp\left[ (\lambda_\mu -\lambda_N) t + 2\int^t dt' \; \frac{3}{8t'} \right]
\simeq
s_\mu(0)   \; e^{ (\lambda_\mu -\lambda_N) t }  \;  t^{3/4} 
\; , 
\end{equation}
which at times of the order of the cross-over to the next regime $t\sim N^{2/3}$ becomes
\begin{equation}
s_\mu(t) \simeq
s_\mu(0)   \; e^{ (\lambda_\mu -\lambda_N) N^{2/3} }  \;  N^{1/2} 
\; . 
\end{equation}
It is clear that the modes which are at a larger distance than $N^{2/3}$ from the edge decay exponentially. 
Instead, those which are closer to $\lambda_N$ than  $N^{2/3}$ grow with the system size. In particular, 
the projection on the  last mode goes as $s_N(t) \simeq s_\mu(0)   \;  N^{1/2} $ by the end of this 
scale, and the mode has already gained a weight that scales with $N$ as in the equilibrium configuration. The 
rest of the evolution, in the second and third regimes, should then serve to make the weight identical to 1, 
as needed in the $T=0$ ground state. 

We finally note that, as long as the initial conditions are of the flat kind, the fluctuations that they induce can be 
neglected and focus only on the fluctuations provoked by the random matrix, as we have done.   
We next turn to the study of the fluctuations induced by special initial conditions, which are near the stationary points of 
the potential energy landscape, the eigenvectors $\vec V_\mu$ of the random interaction matrix. 

\section{Initial states near metastable points}
\label{sec:correlated}

We now fix the random matrix ${\mathbf J}$  and we study the fluctuations with respect to 
 initial conditions which are almost aligned with the 
eigenvectors of this matrix, the projected initial states (ii) defined in Sec.~\ref{subsec:initial}.

\subsection{The escape time-scale}
\label{sec:The escape time-scale}

Starting with an initial condition of the form presented in eq.~(\ref{eq:initial-proj}) we can identify a time scale after which the dynamics are no
longer dominated by the direction of the initial state, i.e.  $\vec V_\alpha$. In order to see this, 
in the equation that fixes $z\big(t,\{s_\mu(0)\}\big)$, eq.~(\ref{eq:fixing-z}), 
we first separate the contribution of  the  $\alpha$ mode  from the one of the rest of the eigenmodes:
\begin{eqnarray}
z\big(t,\{s^{\rm proj}_\mu(0)\}\big)
&=&
\frac{\sum_{\mu(\neq\alpha)} \lambda_\mu \varepsilon^2_\mu\;e^{2\lambda_\mu  t}+\lambda_\alpha (N-\vec{\varepsilon} \cdot \vec{\varepsilon}\,)\;e^{2\lambda_\alpha  t}}{\sum_{\mu(\neq\alpha)} \varepsilon^2_\mu\;e^{2\lambda_\mu  t}+ (N-\vec{\varepsilon} \cdot \vec{\varepsilon}\,)\;e^{2\lambda_\alpha  t}}
\;.
\end{eqnarray}
Next we simplify the analysis by considering that all $\varepsilon_\mu$ are equal to a typical 
value called $\varepsilon_{typ}\ll 1$, see eq.~(\ref{eq;varepsilonmu}), case (ii.a). 
 In the large $N$ limit, $z\big(t,\{s^{\rm proj}_\mu(0)\}\big)$ boils down to
\begin{equation}
\label{eq: annealed calculation}
z\big(t,\{s^{\rm proj}_\mu(0)\}\big) = 
\dfrac{\varepsilon^2_{typ}\displaystyle{\int_{-2J}^{2J}} d\lambda \; \rho(\lambda) \lambda\;e^{2\lambda  t}+\lambda_\alpha (1-\varepsilon_{typ}^2)\;e^{2\lambda_\alpha  t}}{\varepsilon^2_{typ}\displaystyle{\int_{-2J}^{2J}}  d\lambda \; \rho(\lambda) \;e^{2\lambda  t}+ (1-\varepsilon_{typ}^2)\;e^{2\lambda_\alpha  t}}
\end{equation}
which yields an exact expression when $\rho(\lambda)$ is the Wigner semi-circle law
\begin{equation}
\label{eq:z_t}
z\big(t,\{s^{\rm proj}_\mu(0)\}\big) = 
\frac{
\varepsilon^2_{typ}I_2(2\lambda_N t)/t
+
\lambda_\alpha (1- \varepsilon^2_{typ}) \;e^{2\lambda_\alpha  t}}{\varepsilon^2_{typ} I_1(2\lambda_N t)/(\lambda_N t)  + (1-\varepsilon^2_{typ}\,) \; e^{2\lambda_\alpha  t} }
\end{equation}
with  $I_1(y)$ and $I_2(y)$ modified Bessel functions.  At the initial time $t\to 0$, we can use $\lim_{y\to 0} I_a(y) = (y/2)^a/\, \Gamma(a+1)$, and prove that
\begin{equation}
\lim_{t\to 0} 
z\big(t,\{s^{\rm proj}_\mu(0)\}\big)
=
\lambda_\alpha \left(1- \varepsilon^2_{typ}\right) 
\; , 
\end{equation}
while in the long time limit $\lambda_N t \gg 1$, 
$I_a(y) \to e^{y}/\sqrt{2\pi y}$, and 
\begin{eqnarray}
\label{eq: z_annealed_simplified}
\lim\limits_{\lambda_N t\gg1} z\big(t,\{s^{\rm proj}_\mu(0)\}\big)=\frac{\varepsilon_{typ}^2 e^{2\lambda_N t}/(t\sqrt{4\pi\lambda_N t})+\lambda_\alpha (1-\varepsilon_{typ}^2)e^{2\lambda_\alpha t}}{\varepsilon_{typ}^2 e^{2\lambda_N t}/(\lambda_N t\sqrt{4\pi\lambda_N t})+(1-\varepsilon_{typ}^2)e^{2\lambda_\alpha t}}
\, .
\end{eqnarray}
If we assume that the first terms dominate the numerator and denominator,
in the strict infinite time limit
one recovers $\lim_{t\to\infty} z\big(t,\{s^{\rm proj}_\mu(0)\}\big)\to \lambda_N$. 
Otherwise, one can argue that the bulk weights as much as the mode $\alpha$ when the two 
contributions are roughly of the same order. Focusing on the ones in the numerator
\begin{eqnarray}
&&
\varepsilon^2_{typ} \; e ^{2 \lambda_N t}/\sqrt{4\pi \lambda_N t^3}
\sim 
\lambda_\alpha (1-\varepsilon_{typ}^2)\;e^{2\lambda_\alpha  t}
\; . 
\end{eqnarray}
After some rearrangements 
and taking a $\ln$
\begin{eqnarray}
-2(\lambda_N-\lambda_\alpha) t 
\sim
\ln \frac{\varepsilon^2_{typ}}{ (1-\varepsilon_{typ}^2)} - \frac{1}{2} \ln (4\pi) 
- \frac{1}{2} \ln (\lambda^{2}_\alpha \lambda_N t^3)
\; . 
\end{eqnarray}
Assuming that the first term dominates the right-hand-side, in particular,  that $\ln \varepsilon_{typ} \gg \ln(\lambda^{2}_\alpha \lambda_N t^3)$, and 
$\lambda_\alpha \neq \lambda_N$,  
\begin{eqnarray}
t\sim - \frac{1}{\lambda_N-\lambda_\alpha} \, \ln \varepsilon_{typ} 
\equiv t_{\rm esc}
\; , 
\label{eq:time-scale}
\end{eqnarray}
where we defined a time-scale $t_{\rm esc}$. The same time scale would be identified from the analysis of the 
denominator in eq.~(\ref{eq: z_annealed_simplified}).
 As a consequence, if $t$ is larger than the escape time scale outlined in eq.~(\ref{eq:time-scale}), the dynamics describe the departure of the system from its 
 initial state  $\vec{S}(0)\approx\sqrt{N}\vec{V}_\alpha$. We will see from the numerical solution of the dynamic equations, averaged over 
 many such initial conditions, that this time scale characterises the cross over of  $z\big(t,\{s^{\rm proj}_\mu(0)\}\big)$ 
 from $\lambda_\alpha$ to $\lambda_N$. We point out that this analysis is naive in the sense that we have assumed that the 
 bulk contribution can be replaced by an average over Wigner's semi-circle law 
 and the dependence on the initial conditions can be grasped from  one typical realisation. 
 Note that taking $\varepsilon_{typ}\sim N^{-\nu}$
  the time-scale discussed above is 
 \begin{equation}
 \varepsilon_{typ}\sim N^{-\nu}
 \qquad\quad
 \implies 
  \qquad\quad
 t_{\rm esc}\sim  \frac{\nu}{\lambda_N-\lambda_\alpha} \, \ln N 
 \label{eq:t-esc}
 \end{equation}
  and, therefore, quite short. The escape from the initial condition occurs quickly, though in a time 
 scale that grows with $N$. 
 
\subsection{Loss of self-averageness}
\label{sec:loss}

In the previous Section we obtained a naive expression for $z$, eq.~(\ref{eq:z_t}), 
that separates  the contribution of the  mode $\alpha$, along which the 
 initial state is mostly aligned, from the one of the rest of the bulk. From it we identified the $N$-dependent 
 escape time $t_{\rm esc} \sim \ln N$ for $\varepsilon_\mu = \varepsilon_{typ} = N^{-\nu}$ $\forall \mu$.
 In the following, we  present an evaluation of $z$ 
 where we carefully take the average over the projected initial conditions. It   indicates that at this time-scale 
strong fluctuations with respect to the initial conditions appear.

\subsubsection{The partition function point of view}
 
As already mentioned in eq.~(\ref{eq:fixing-z}), for all initial conditions and any $N$, 
the integral over time of the Lagrange multiplier reads
\begin{eqnarray}
2\int_{0}^{t}dt'\, z(t',\{s_\mu(0)\})=\ln\Big[\frac{1}{N}\sum_\mu  s^2_\mu(0)\, e^{2\lambda_\mu t}\Big]=\ln\Big[{\mathcal Z}_N\big(t,\{s_\mu(0)\}\big)\Big]\, .
\label{eq:Gamma}
\end{eqnarray}
The function ${\mathcal Z}_N\big(t,\{s_\mu(0)\}\big)$ can be interpreted as a partition function in the usual thermodynamics context.\footnote{It is the 
function called $\Gamma$ in~\cite{CuDe95a,CuDe95b}.} Here, the time $t$ plays the role of the inverse temperature 
$\beta$, $-2\lambda_\mu$ the role of the energy $E_\mu$, and $s^2_\mu(0)$ the role of a degeneracy $\Omega\big(E_\mu\big)$. 
Therefore, as time evolves the system is annealed in the sense of following  
 lower and lower temperatures $T=1/\beta=1/t$. In the infinite time limit the system falls in the ground state 
 corresponding to the lowest energy density $e_{\rm ground}=-\lim_{t\to\infty}z/2= -\lambda_N/2$.
 The correspondence with thermodynamics goes even further as the outlined partition function enables one to determine 
most of the relevant observables. As an example we have 
\begin{eqnarray}
z(t,\{s_\mu(0)\})=-2e[\vec{S}(t)]=\frac{1}{2}\frac{d}{dt} \ln\Big[{\mathcal Z}_N\big(t,\{s_\mu(0)\}\big)\Big]
\; . 
\end{eqnarray}
In the following we will determine the value taken by $z\big(t,\{s_\mu(0)\}\big)$ when the initial state $\vec{S}(0)$ is 
averaged out, indicated by the notation $\langle \dots \rangle_{i.c.}$.
We will see that for projected initial conditions, beyond the time-scale $t_{\rm esc}$ the dynamics will not ``self-average'' with respect
to the initial configurations.

To exactly evaluate $\langle\ln[{\mathcal Z}_N(t,\{s_\mu(0)\})]\rangle_{i.c.}$ we separate the contributions from the bulk and the 
$\alpha$ mode:
\begin{eqnarray}
\ln\Big[{\mathcal Z}_N\big(t,\{s^{\rm proj}_\mu(0)\}\big)\Big]=\ln\Big[\frac{1}{N}\sum_\mu \ (s^{\rm proj}_\mu(0))^2 
\ e^{2\lambda_\mu t}\Big]=\ln\Big[\mathcal{Z}_{\rm bulk}\big(t,\{s^{\rm proj}_\mu(0)\}\big)+\frac{1}{N} (s^{\rm proj}_\alpha(0))^2  \, e^{2\lambda_\alpha t}\Big]
\end{eqnarray}
with
\begin{eqnarray}
\mathcal{Z}_{\rm bulk}\big(t,\{s^{\rm proj}_\mu(0)\}\big)=\frac{1}{N}\sum_{\mu(\neq\alpha)} 
(s^{{\rm proj}}_\mu(0))^2 \ e^{2\lambda_\mu t}
\; .
\end{eqnarray}
We recall that the projected initial conditions verify 
\begin{eqnarray}
s^{\rm proj}_{\mu(\neq \alpha)}(0)
= X
\qquad
\mbox{and}
\qquad
s^{\rm proj}_{\alpha}(0)= \sqrt{N-X^2 (N-1)}\approx\sqrt{N}
\end{eqnarray}
where $X$ is a Gaussian random variable with zero mean $\langle X\rangle_{i.c.}=0$ and $\langle X^2\rangle_{i.c.} =\varepsilon^2_{typ} \ll 1$, 
properties that ensure 
$\varepsilon^2 \ll N$. We can thus rewrite the logarithm of the partition function as
\begin{eqnarray}
&& \ln\Big[ {\mathcal Z}_N\big( t,\{s^{\rm proj}_\mu(0)\} \big)\Big]
=
\ln\Big[\frac{X^2}{N}\sum_{\mu(\neq\alpha)}e^{2\lambda_\mu t }+e^{2\lambda_\alpha t }\Big]
=\ln\Big[X^2\mathcal{Y}_{\rm bulk}(t)+e^{2\lambda_\alpha t}\Big]
\;\;\;\;
\end{eqnarray}
with
\begin{eqnarray}
\mathcal{Y}_{\rm bulk}(t)= 
\frac{1}{N}\sum_{\mu(\neq\alpha)}e^{2\lambda_\mu t}=\int d\lambda\, \rho(\lambda)\,e^{2\lambda t}
=
\frac{I_1(2\lambda_N t)}{\lambda_N t} 
\underset{t\to\infty}{\longrightarrow} \frac{e^{2\lambda_N t}}{\sqrt{4\pi} (\lambda_N t)^{3/2}}
\; .
\end{eqnarray}
The next subsection detail
how to study the fluctuations of 
$\ln\Big[{\mathcal Z}_N\big(t,\{s^{\rm proj}_\mu(0)\} \big)\Big]$
induced by these random initial conditions.

\subsubsection{The distribution of ${\mathcal Z}_N\big( t,\{s^{\rm proj}_\mu(0)\} \big)$}

The probability distribution $\rho\big[{\mathcal Z}_N\big( t,\{s^{\rm proj}_\mu(0)\} \big)\big]$, and its evolution with time,  
can be derived from the probability distribution of the initial 
configurations $\{s_\mu^{\rm proj}(0)\}$:
\begin{eqnarray}
\rho\big[{\mathcal Z}_N\big( t,\{s^{\rm proj}_\mu(0)\} \big)\big]
&=&
\int \frac{dk}{2\pi}\frac{dX}{\sqrt{2\pi \varepsilon_{typ}^2}} 
\; 
\exp\Bigg[\frac{-X^2}{2\varepsilon_{typ}^2}+ik\Big({\mathcal Z}_N\big( t,\{s^{\rm proj}_\mu(0)\}\big)-X²\mathcal{Y}_{\rm bulk}(t)-e^{2\lambda_\alpha t}\Big)\Bigg]
\; ,
\end{eqnarray}
where we used the Fourier representation of the Dirac delta.
Integrating out the variable $X$ we obtain
\begin{eqnarray}
\label{eq: exact distribution}
\rho\big[{\mathcal Z}_N\big( t,\{s^{\rm proj}_\mu(0)\} \big)\big]
&=&\int \frac{dk}{2\pi}\exp\Bigg[ik{\mathcal Z}_N\big( t,\{s^{\rm proj}_\mu(0)\} \big)-ike^{2\lambda_\alpha t}-\frac{1}{2}\ln\Big(1+2ik\varepsilon_{typ}^2\, \mathcal{Y}_{\rm bulk}(t) \Big)\Bigg]\nonumber\\
&\equiv& \int \frac{dk}{2\pi}\exp\Big[S\big(k,{\mathcal Z}_N\big( t,\{s^{\rm proj}_\mu(0)\} \big)\Big]\;.
\end{eqnarray}
A first approach to simplify the action $S\big(k,{\mathcal Z}_N\big( t,\{s^{\rm proj}_\mu(0)\} \big)$ is to Taylor expand it around $k=0$:
\begin{eqnarray}
\label{eq: exact distribution series}
\rho\big[{\mathcal Z}_N\big( t,\{s^{\rm proj}_\mu(0)\} \big)\big]
&=&\int \frac{dk}{2\pi}\exp\Bigg[ik{\mathcal Z}_N\big( t,\{s^{\rm proj}_\mu(0)\} \big)-ike^{2\lambda_\alpha t}+\frac{1}{2}\sum_m \frac{\big(-2ik\varepsilon_{typ}^2 \,  \mathcal{Y}_{\rm bulk}(t)\big)^{m}}{m}\Bigg]\;
\end{eqnarray}
with the condition $2k\varepsilon_{typ}^2 \,  \mathcal{Y}_{\rm bulk}(t)<1$. By limiting the Taylor expansion to second order in $k$ it follows that
\begin{eqnarray}
\rho\big[{\mathcal Z}_N\big( t,\{s^{\rm proj}_\mu(0)\} \big)\big]
&\sim &\int \frac{dk}{2\pi}\exp\Bigg[ik{\mathcal Z}_N\big( t,\{s^{\rm proj}_\mu(0)\} \big)-ike^{2\lambda_\alpha t}-ik\varepsilon_{typ}^2 \,  \mathcal{Y}_{\rm bulk}(t)-k^2 \varepsilon_{typ}^4 \,\mathcal{Y}^2_{\rm bulk}(t)\Bigg]\nonumber\\
&\sim&\frac{1}{\mathcal{N}'}
\exp\Bigg[\Big(
{\mathcal Z}_N\big( t,\{s^{\rm proj}_\mu(0)\} \big)
-e^{2\lambda_\alpha t}-\varepsilon_{typ}^2 \,  \mathcal{Y}_{\rm bulk}(t)\Big)^2/\Big(4\,\varepsilon_{typ}^4\, \mathcal{Y}^2_{\rm bulk}(t)\Big)\Bigg]
\end{eqnarray}
with
$
\mathcal{N}'=\sqrt{4\pi\,\varepsilon_{typ}^4 \,\mathcal{Y}^2_{\rm bulk}(t)}
$.
To this order, the distribution of ${\mathcal Z}_N\big( t,\{s^{\rm proj}_\mu(0)\} \big)$ is therefore Gaussian with mean 
\begin{equation}
\Big\langle {\mathcal Z}_N\big( t,\{s^{\rm proj}_\mu(0)\} \big) \Big\rangle_{i. c.} 
= 
e^{2\lambda_\alpha t}+\varepsilon_{typ}^2 \,  \mathcal{Y}_{\rm bulk}(t)
 \underset{N\rightarrow\infty}{=} 
e^{2\lambda_\alpha t} + \varepsilon_{typ}^2  \ \frac{I_1(2\lambda_N t)}{\lambda_N t}
  \underset{t\rightarrow\infty}{=}  
  e^{2\lambda_\alpha t}+\frac{\varepsilon_{typ}^2}{(2\pi)^{1/2}}  \  \frac{e^{2\lambda_N t}}{ (\lambda_N t)^{3/2}}
\end{equation}
 and variance 
\begin{eqnarray}
\Big\langle 
\left(
 {\mathcal Z}_N\big( t,\{s^{\rm proj}_\mu(0)\}\big) \! - \! \Big\langle  {\mathcal Z}_N\big( t,\{s^{\rm proj}_\mu(0)\}\big) \Big\rangle_{i. c.}
\right)^2
\Big\rangle_{i. c.}
\!\!\!\!\!\!\!
= 
2\,\varepsilon_{typ}^4\, \mathcal{Y}^2_{\rm bulk}(t)
 \underset{N\rightarrow\infty}{\longrightarrow} 
2 \,\varepsilon_{typ}^4 \Big( \frac{I_1(2\lambda_N t)}{\lambda_N t}\Big)^2
 \!\!\! \underset{t\rightarrow\infty}{\longrightarrow}  \!
\frac{\varepsilon_{typ}^4}{\pi}   \frac{e^{4\lambda_N t}}{(\lambda_N t)^{3}} 
\; . 
\end{eqnarray} 

As in Sec.~\ref{sec:The escape time-scale}, the characteristic time-scale $t_{\rm esc}$, see eq.~(\ref{eq:time-scale}), 
distinguishes two dynamic regimes  in the large $N$ limit. 
When $t\ll t_{\rm esc}$, the standard deviation of the Gaussian distribution is negligible compared to its mean value. 
In this case the density distribution collapses to a delta function, i.e.
\begin{eqnarray}
 \rho\big[{\mathcal Z}_N\big( t,\{s^{\rm proj}_\mu(0)\} \big)\big]
 =\delta\Big({\mathcal Z}_N\big( t,\{s^{\rm proj}_\mu(0)\} \big)-e^{2\lambda_\alpha t}-\varepsilon_{typ}^2 \,  \mathcal{Y}_{\rm bulk}(t)\Big)
\end{eqnarray}
 and, straightforwardly, we recover the results of the  previous Subsection.
For instance, 
\begin{eqnarray}
2\Big\langle\int_{0}^{t}dt'\, z(t',\{s^{\rm proj}_\mu(0)\})\Big\rangle_{i.c.}=\Big\langle\ln\Big[{\mathcal Z}_N\big(t,\{s^{\rm proj}_\mu(0)\}\big)\Big]\Big\rangle_{i.c.}=\ln\Big[e^{2\lambda_\alpha t}+\varepsilon_{typ}^2 \,  \mathcal{Y}_{\rm bulk}(t)\Big]
\label{eq: simplify1}
\end{eqnarray}
and thus
\begin{eqnarray}
\Big\langle z\big(t,\{s^{\rm proj}_\mu(0)\}\big)
\Big\rangle_{i.c.}=
\frac{
\varepsilon^2_{typ}I_2(2\lambda_N t)/t
+
\lambda_\alpha \;e^{2\lambda_\alpha  t}}{\varepsilon^2_{typ} I_1(2\lambda_N t)/(\lambda_N t)  + \; e^{2\lambda_\alpha  t} }
\; ,
\end{eqnarray}
which is identical to eq.~(\ref{eq:z_t}).

However, when $t\gg t_{\rm esc}$, the mean and standard deviation of the Gaussian distribution scale evenly with the system size and the 
distribution does not simplify to a delta function. Besides, in this regime the truncation of the series in eq.~(\ref{eq: exact distribution series}) 
up to second order in $k$, and more generally the Taylor expansion of the logarithm, are no longer justified. Indeed, for a non-negligible part of the integral we verify $k\,\varepsilon_{typ}^2\, \mathcal{Y}_{bulk}(t)=\mathcal{O}(1)$, making all terms in the series count. The distribution of ${\mathcal Z}_N\big( t,\{s^{\rm proj}_\mu(0)\} \big)$ now exhibits fluctuations, meaning that the dynamics depend non-negligibly on its initial conditions. 

In App.~\ref{app:exact} we detail an alternative method to compute
$\Big\langle\ln\Big[{\mathcal Z}_N\big(t,\{s^{\rm proj}_\mu(0)\}\big)\Big]\Big\rangle_{i.c.}$ in an exact manner, 
which has the benefit of also being more tractable numerically.

\subsection{Numerical solution}

To keep track of the fluctuations caused by the distribution of $\vec{\varepsilon}$, 
all our results are obtained with one fixed interaction matrix $\mathbf {J}$ with 
size $N\times N$ and averaged over 500 realizations of $\vec{\varepsilon}$. 
We used $\varepsilon_{typ}=N^{-1/2}$, i.e. $\nu=1/2$.
In the following we represent the average over initial conditions with the notation 
$\langle \dots \rangle_{i.c.}$ and its standard deviation with $\Delta (\dots)$. As an example we consider
\begin{eqnarray}
\Delta z\big(t,\{s_\mu(0)\}\big)=
\langle z²\big(t,\{s_\mu(0)\}\big)\rangle_{i.c.}-\langle z\big(t,\{s_\mu(0)\}\big)\rangle²_{i.c.} 
\; .
\label{eq:fluct-z}
\end{eqnarray}

\begin{figure}[h!]
\hspace{6cm} (a) \hspace{7cm} (b)
\vspace{-0.5cm}
    \begin{center}
    \includegraphics[width=0.45\textwidth]{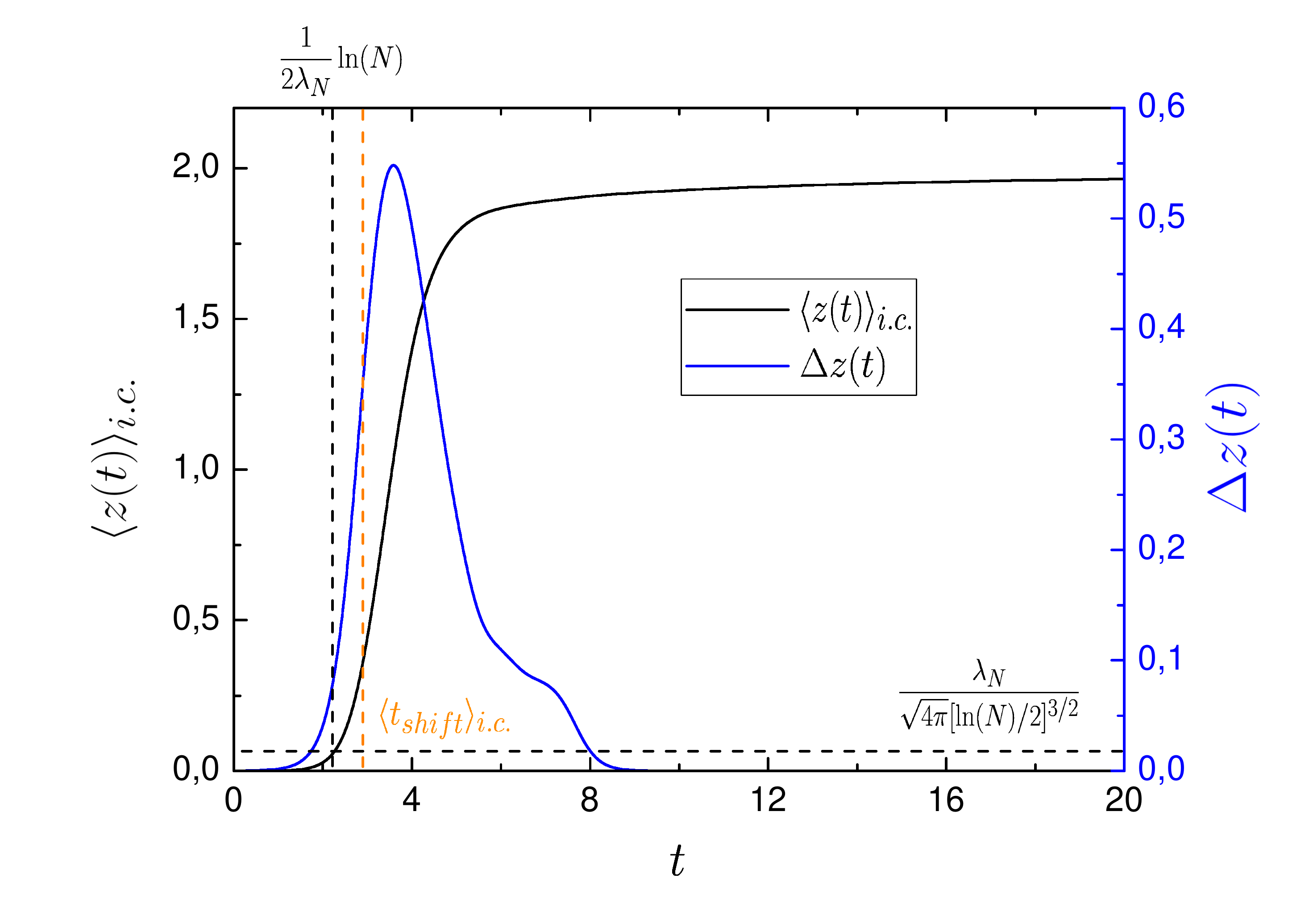}
   \includegraphics[width=0.45\textwidth]{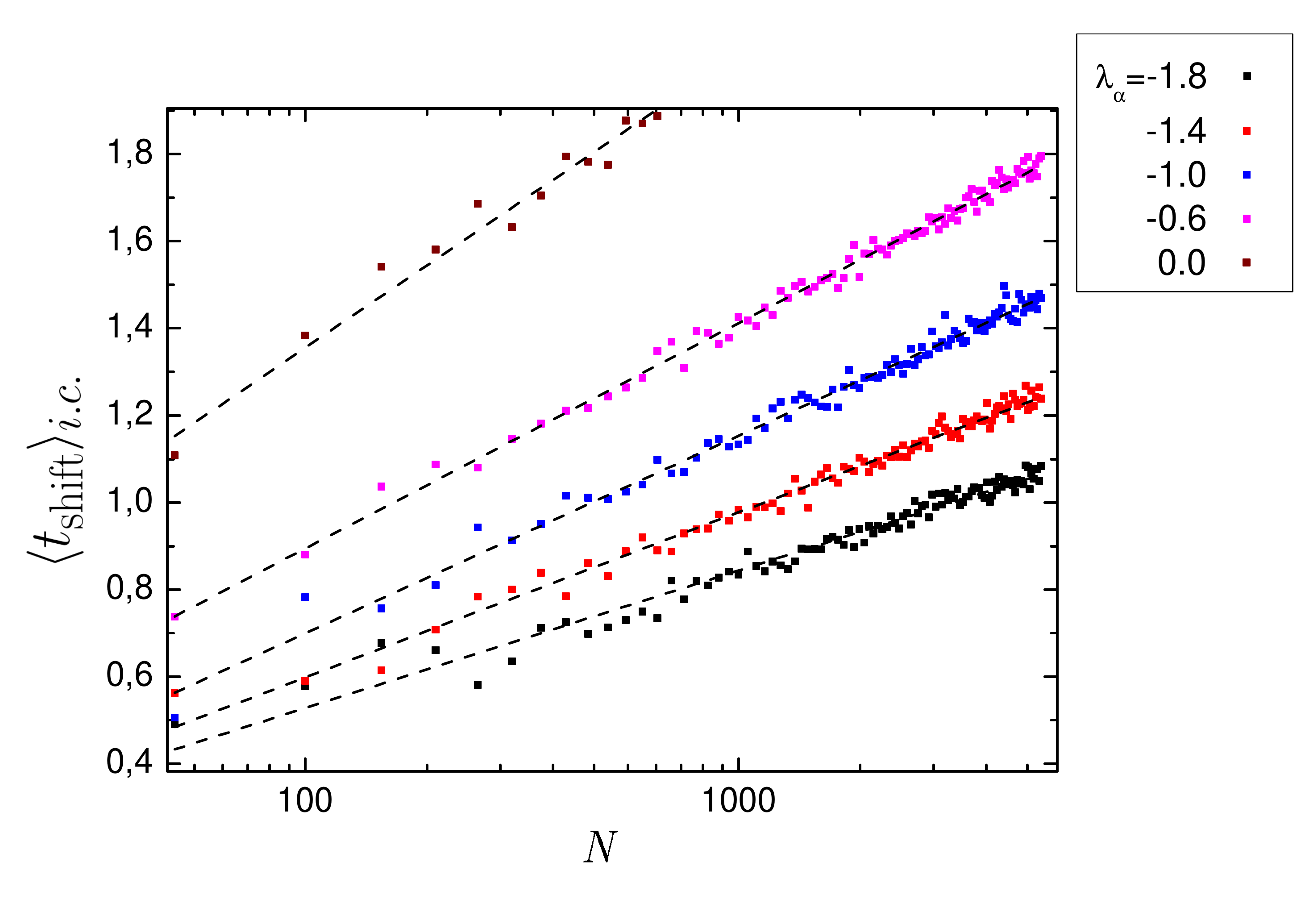}
        \end{center}
        \vspace{-0.5cm}
    \caption{\small (a)     Time evolution of the Lagrange multiplier in a model with
    $N=5000$ with  $\lambda_\alpha \sim 0$, and 
    sampling data over 500 realisations of $\vec{\varepsilon}$ all with $z(0) \sim 0$ and randomness as in (ii).
     The  black solid curve is the result of the microscopic dynamics averaged over  initial conditions.
    The blue solid curve shows the fluctuations of $z$, $\Delta z$, as defined in eq.~(\ref{eq:fluct-z}).  
    The vertical dashed black line is the escape time (\ref{eq:t-esc}) for 
    $\lambda_\alpha=0$ and the horizontal one $z(t_{\rm esc})$  
    The vertical dashed orange line is the averaged value of $t_{\rm shift}$, the 
    time when $z\big(t,\{s^{\rm proj}_\mu(0)\}\big)$  becomes convex.   
         (b) Evolution of $t_{\rm shift}$ with  the number of spins $N$. 
              The dashed curves are guides-to-the-eye of the form $a_\alpha+b_\alpha\ln N$. 
              Although it verifies $b_\alpha\propto\frac{1}{\lambda_N-\lambda_\alpha}$ the data deviate a bit from the other pre-factor $\nu$ 
              expected from Eq.~(\ref{eq:t-esc}).
     }
    \label{fig:z_variation}
\end{figure}

The time scale $t_{\rm esc}$ in eq.~(\ref{eq:time-scale}) corresponds to the escape time  
from the initial state.
For a system with $N=5000$ and $\lambda_\alpha=0$ it equals $t\sim 1/(2\lambda_N) \ln N \simeq 
2.13$.
In Fig.~\ref{fig:z_variation}(a) we observe the departure from the initial value, 
with the sudden shift of the 
Lagrange multiplier $z\big(t,\{s^{\rm proj}_\mu(0)\}\big)$, changing from $z=\lambda_\alpha=0J$ to a function converging to 
$\lambda_N=2J$. Numerically, we can identify a departure time 
$t_{\rm shift}$ as the time at which the function $z$ becomes concave,  
\begin{equation}
\label{eq: tshift def}
    \partial_t^2 z\big(t,\{s_\mu(0)\}\big)\Big|_{t_{\rm shift}}= \; 0 \; .
\end{equation} 
Concretely, we calculate $t_{\rm shift}$ for each run and we then average over all of them to 
obtain $\langle t_{\rm shift}\rangle_{i.c.}$.
$t_{\rm shift}$ is not necessarily equal to the time
scale we identified with analytic arguments, $t_{\rm esc}$, but one can expect it to be of 
the same order of magnitude and to scale with $N$, and depend on other parameters, in a similar way. 
This is confirmed by the numerical results shown in Fig.~\ref{fig:z_variation}(a).
In eq.~(\ref{eq:time-scale}) we pointed out that the characteristic time-scale should scale with $N$ as ${\ln{N}}$, 
when we take $\varepsilon_{typ}\sim 1/\sqrt{N}$. Consequently, in Fig.~\ref{fig:z_variation}(b)
we plotted $t_{\rm shift}$ as function of $N$ and we obtained very good agreement with this prediction. 

\begin{figure}[h!]
\vspace{0.25cm}
    \hspace{0cm} (a)    \hspace{7cm} (b)
    \vspace{-0.25cm}
    \\
        \centering
             \includegraphics[width=0.45\textwidth]{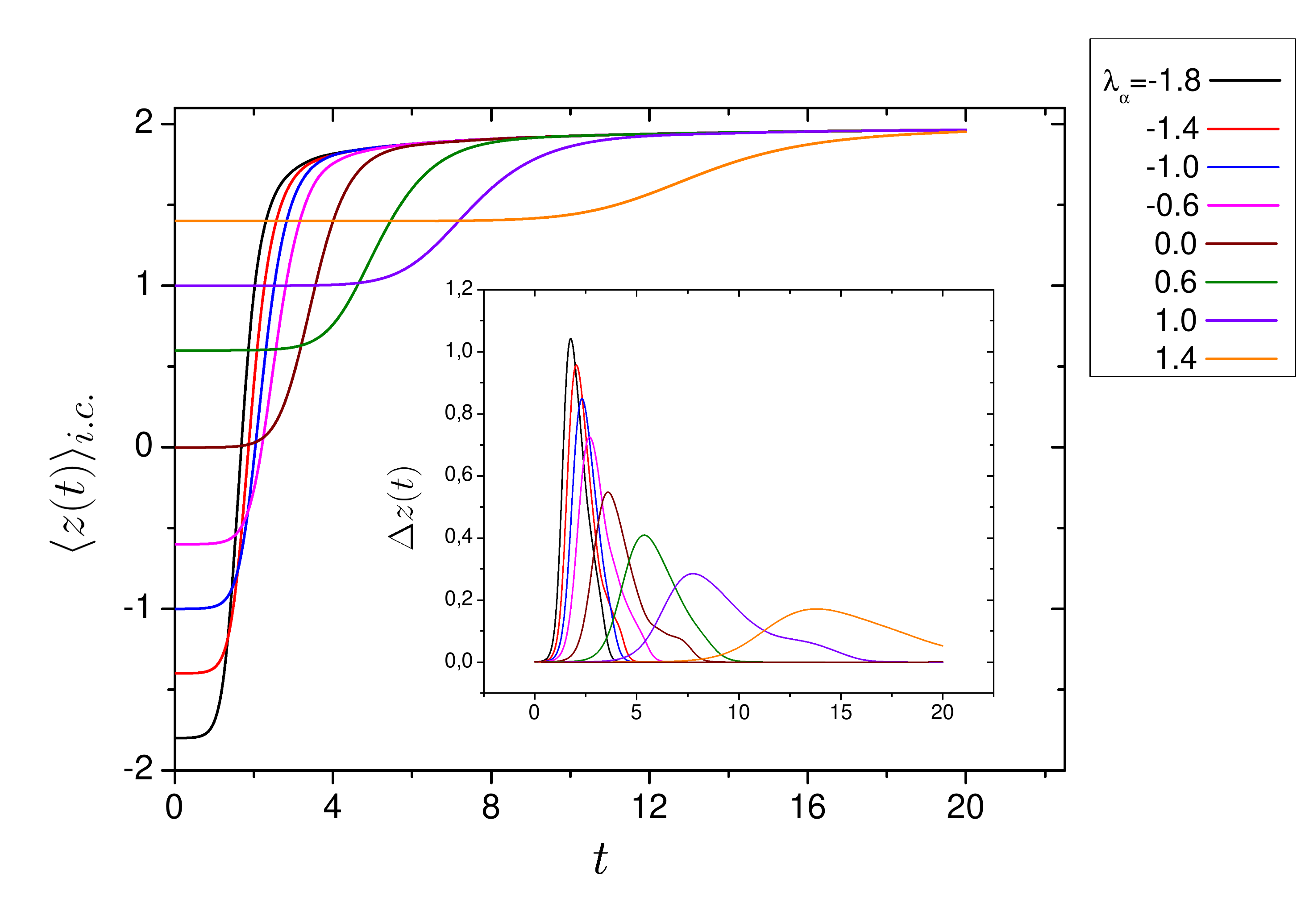}
        \includegraphics[width=0.45\textwidth]{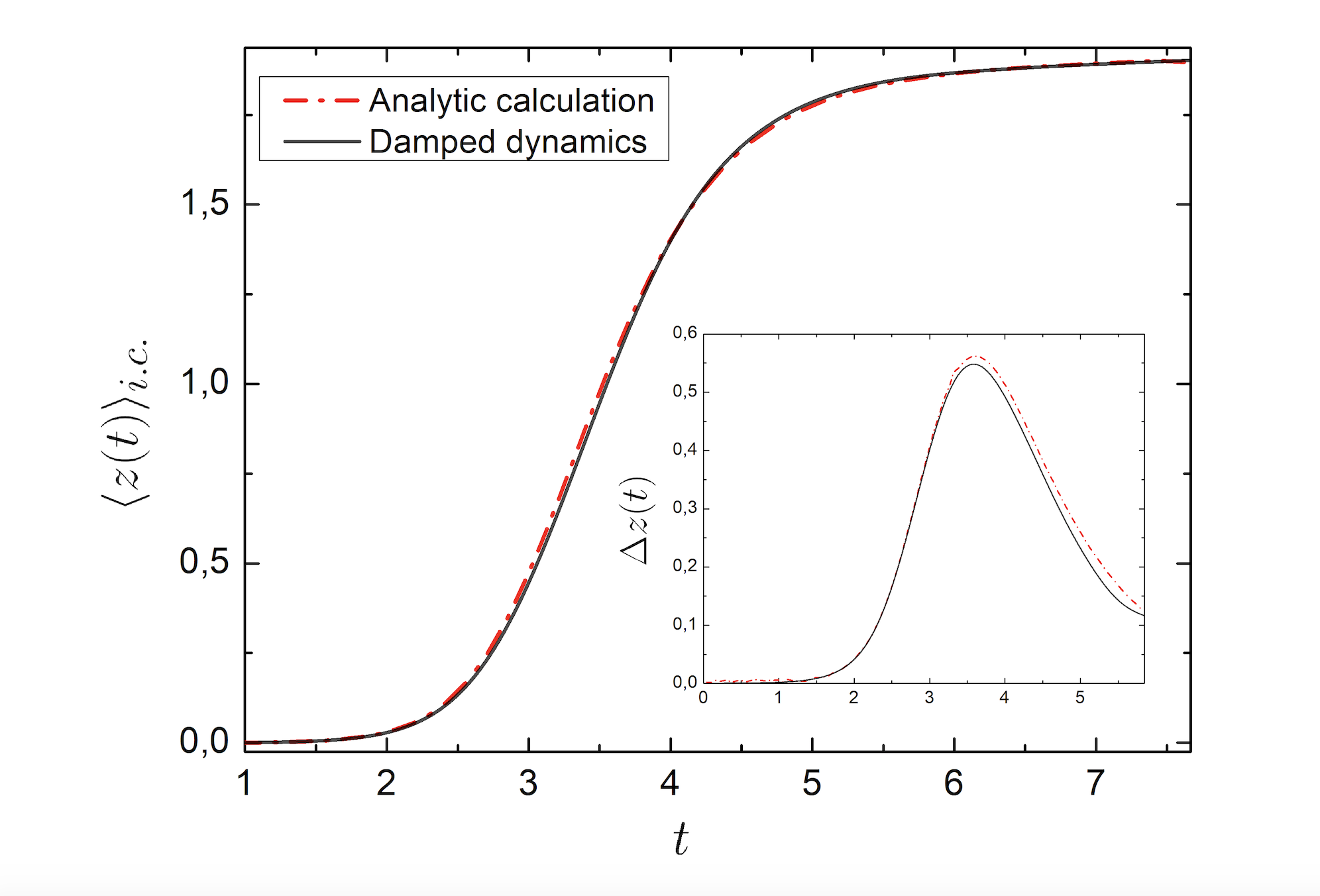}
  \caption{\small 
 (a) Time evolution of  the averaged Lagrange multiplier after 
 being averaged over projected initial conditions with different choices of $\vec V_\alpha$. 
 In the inset, the fluctuations $\Delta z\big(t,\{s^{\rm proj}_\mu(0)\}\big)$ for the same parameters.
  (b) Comparison to the analytic predictions obtained 
  with the free-energy $\langle \ln[{\mathcal Z}_N\big(t,\{s^{\rm proj}_\mu(0)\}\big)] \rangle_{i. c.}$ 
  for $\lambda_\alpha=0$.
  In the main plot the average of $z$ and in the inset its fluctuatins.
  In both plots the system size is $N=5000$ and 500 realisations of the 
  initial condition are used.
   }
      \label{fig:energy_variation}
\end{figure}

Figure~\ref{fig:energy_variation} shows similar results for various initial states, $\vec S(0) = \vec V_\alpha + \vec \varepsilon$. In all cases the 
averaged Lagrange multiplier approaches the asymptotic value $\lambda_N$, but the pre-factor of 
time scales $t_{\rm esc}$ and $t_{\rm shift}$ depends on $\lambda_\alpha$. In the inset we display the fluctuations of 
$z$ for the same choices of initial conditions. Panel (b) in the same figure confronts the numerical results to the 
analytic predictions developed in App.~A, from variations of eq.~(\ref{eq:replicas}). The agreement is excellent for 
the averaged value and there are some tiny deviations for the fluctuations that diminish for increasing system size.

\section{Discussion}
\label{sec:conclusions}

In this paper we revisited the relaxation dynamics of the spherical Sherrington-Kirkpatrick or $p=2$ model.
For infinite temperature initial conditions, we verified the predictions made in Refs.~\cite{CuDe95a} and \cite{FyPeSc15}
for two algebraically decaying regimes of the excess energy density. In the first one, time is finite with respect to $N$,
and the relaxation is controlled by the decay of $\rho(\lambda)$ at its edge in the $N\to\infty$ limit.
The evolution 
crosses over to a second regime at a time-scale that depends on $N$ in a way determined by the distribution of 
the first gap, $\rho_{\rm gap}(g_N)$. For GOE interaction matrices  $t_{\rm cross} \sim N^{2/3}$.
In terms of the time dependencies, e.g the excess energy relaxation in the second algebraic
  regime is faster than in the first one, $t^{-3}$ vs. $t^{-1}$.
We employed much larger systems than 
previously used~\cite{FyPeSc15} and our numerical data fully confirm these predictions. 
We also identified the crossover to a last time 
regime in which the minimum first gap, $g^{\rm min}_N$, in a random matrix sampling of finite size, 
controls the average and is the inverse time-scale of the final exponential decay.
For initial conditions mostly aligned with eigenvectors of the random exchange matrix, we found a 
time scale $t_{\rm esc} \sim t_{\rm shift} \sim \ln N$ beyond which the system departs considerably from the initial state and 
the dynamics lose their ``self-averageness'' with respect to the initial conditions. We proved this fact
{\it via} a mapping to an effective partition function in which time plays the role of the inverse temperature. We calculated 
the corresponding free-energy with the replica method and from it we derived the 
average energy density and its fluctuations finding again excellent agreement with the 
numerical evaluation of the same quantities. 

This model, albeit relatively simple since almost quadratic in nature, 
finds applications in many branches of physics and its interfaces. Some very recent examples are 
neural networks~\cite{MaBrOs18,FrScUr19},  signal encryption~\cite{Fy19}, 
chaos in the classical limit of the Sachdev-Ye-Kitaev model~\cite{ScAl19}, and
integrability~\cite{BaCuLoNe20}. Fully understanding its behaviour for finite number of degrees of 
freedom will be of help in these areas as well.

\vspace{1cm}

\noindent
{\bf \large Acknowldegements.}
We warmly thank G. Schehr for very helpful discussions. D.A.S aknowledges brazilian funding agency CNPq for
partial financial support. 

\vspace{0.25cm}

\appendix

\section{Exact calculation of $\big\langle\ln[{\mathcal Z}_N\big(t,\{s^{\rm proj}_\mu(0)\}\big)] \,\big\rangle_{i.c.}$}
\label{app:exact}


We revisit the calculation of the average of ${\mathcal Z}_N$ over projected initial conditions.
To exactly average over these initial states we apply the replica technique
\begin{eqnarray}
\Big\langle \ln\Big[{\mathcal Z}_N\big(t,\{s^{\rm proj}_\mu(0)\}\big)\Big] \Big\rangle_{i. c.}
\underset{n\rightarrow0}{=}\frac{\Big\langle\Big[X²\mathcal{Y}_{\rm bulk}(t,\{J_\mu\})+e^{2\lambda_\alpha t}\Big]^n\Big\rangle_{i. c.}-1}{n}\;.
\end{eqnarray}
Focusing on the right-hand side of the previous equation we have
\begin{eqnarray}
&& 
\Big\langle\Big[X² \, \mathcal{Y}_{\rm bulk}(t)+e^{2\lambda_\alpha t}\Big]^n \Big\rangle_{i. c.}
=
\sum_{n_\alpha=0}^{n} \ \binom{n}{n_\alpha} \ e^{2\lambda_\alpha n_\alpha t} \ \big\langle X^{2(n-n_\alpha)}\big\rangle_{i. c.} 
\ \mathcal{Y}^{(n-n_\alpha)}_{\rm bulk}(t)
\nonumber\\
&& 
\qquad\quad
=\sum_{n_\alpha=0}^{n}\binom{n}{n_\alpha} \ e^{2\lambda_\alpha n_\alpha t} \
\frac{[2(n-n_\alpha)]!}{(n-n_\alpha)! \,2^{(n-n_\alpha)}} \ \big\langle X^{2}\big\rangle^{n-n_\alpha}_{i. c.}  \ \mathcal{Y}^{(n-n_\alpha)}_{\rm bulk}(t)
\nonumber\\
&&
\qquad\quad
=
\sum_{n_\alpha=0}^{n} \binom{n}{n_\alpha} \ e^{2\lambda_\alpha n_\alpha t} \ \binom{2n-2n_\alpha}{n-n_\alpha} (n-n_\alpha)!\, 
\Bigg(\frac{\big\langle X^{2}\big\rangle_{i. c.} \ \mathcal{Y}_{\rm bulk}(t)}{2}\Bigg)^{(n-n_\alpha)}
\nonumber\\
&&
\qquad\quad
=
\sum_{n_\alpha=0}^{n} \ \frac{e^{2\lambda_\alpha n_\alpha t}}{4\pi²}
\  
\int_{-\pi}^{\pi}dx\int_{-\pi}^{\pi}dx' \, \big(1+e^{ix}\big)^{n}e^{-ix n_\alpha}\big(1+e^{ix'}\big)^{2(n-n_\alpha)}e^{-ix'(n-n_\alpha)}
\nonumber\\
&&
\qquad\quad\hspace{3cm}\times \ \int_{0}^{+\infty}dy \ e^{-y} \ y^{n-n_\alpha}\, \Bigg(\frac{\big\langle X^{2}\big\rangle_{i. c.}\mathcal{Y}_{\rm bulk}(t)}{2}\Bigg)^{(n-n_\alpha)}
\nonumber\\
&&
\qquad\quad
=\frac{1}{4\pi^2}\int_{-\pi}^{\pi}dx\int_{-\pi}^{\pi}dx'\int_{0}^{\infty}dy \ e^{-y} \ \Big[ y \big(1+e^{ix}\big)\big(1+e^{ix'}\big)^{2}e^{-ix'} \Big]^n 
\nonumber\\
\qquad\quad
&&\hspace{4cm}\times \ \Bigg(\frac{\big\langle X^{2}\big\rangle_{i. c.}\mathcal{Y}_{\rm bulk}(t)}{2}\Bigg)^{n} 
\times \ \sum_{n_\alpha=0}^{n}\Bigg(\frac{2\,e^{2\lambda_\alpha t+i(x'-x)}}{y(1+e^{ix'})^2\big\langle X^{2}\big\rangle_{i. c.}
\ \mathcal{Y}_{\rm bulk}(t)}\Bigg)^{n_\alpha} \; .
\nonumber
\end{eqnarray}
Finally if we define the complex function
\begin{eqnarray}
f(t,x,x',y)=\frac{2\,e^{2\lambda_\alpha t+i(x'-x)}}{y(1+e^{ix'})^2\big\langle X^{2}\big\rangle_{i. c.}\mathcal{Y}_{\rm bulk}(t)}
\; , 
\end{eqnarray}
which contains the dependence on the initial condition through $\lambda_\alpha$, 
we obtain
\begin{eqnarray}
\Big\langle\Big[X² \ \mathcal{Y}_{\rm bulk}(t)+e^{2\lambda_\alpha t}\Big]^n
\Big\rangle_{i. c.}&=&\frac{1}{4\pi^2}\int_{-\pi}^{\pi}dx\int_{-\pi}^{\pi}dx'\int_{0}^{+\infty}dy\,e^{-y} \ \Big[ y \big(1+e^{ix}\big)\big(1+e^{ix'}\big)^{2}e^{-ix'} \Big]^n \\
&&
\qquad
\times \ \Bigg(\frac{\big\langle X^{2}\big\rangle_{i. c.}\mathcal{Y}_{\rm bulk}(t)}{2}\Bigg)^{n}\frac{1-f(t,x,x',y)^{n+1}}{1-f(t,x,x',y)} \; .\nonumber
\end{eqnarray}
Taking the limit $n\rightarrow 0$ we determine the logarithm of the partition function, it yields
\begin{eqnarray}
\Big\langle \ln\Big[{\mathcal Z}_N\big(t,\{s_\mu(0)\}\big)\Big] \Big\rangle_{i. c.}&=&\ln\Bigg[\frac{\big\langle X^{2}\big\rangle_{i. c.}\mathcal{Y}_{\rm bulk}(t)}{2}\Bigg]+\int_{0}^{+\infty}dy\, e^{-y}\ln y\nonumber\\
&&-\frac{1}{4\pi^2}\int_{-\pi}^{\pi}dx\int_{-\pi}^{\pi}dx'\int_{0}^{+\infty}dy\,e^{-y}\Bigg[\frac{f(t,x,x',y)\ln[f(t,x,x',y)]}{1-f(t,x,x',y)}\Bigg]\; .
\label{eq:replicas}
\end{eqnarray}
This expression demonstrates that $\langle \ln {\mathcal Z}_N\rangle_{i.c.} \neq \ln \langle {\mathcal Z}_N\rangle_{i.c.}$
and that there is no self-averageness with respect to the initial conditions. We will use it in the main text to 
evaluate the statistical properties of the Lagrange multiplier $z$ and hence of the energy density $e$.

We can readily check the two time limits $t\rightarrow0$ and $t\rightarrow+\infty$.
In the first one as we have  $\langle X²\rangle_{i.c.}\ll 1$, the fraction in eq.~(\ref{eq:replicas}) simplifies to
\begin{eqnarray}
\frac{f(t,x,x',y)\ln[f(t,x,x',y)]}{1-f(t,x,x',y)}\underset{t\rightarrow0}{=}-\ln[f(t ,x,x',y)]
\end{eqnarray}
and we are left with 
\begin{eqnarray}
\Big\langle \ln\Big[{\mathcal Z}_N\big(t,\{s_\mu(0)\}\big)\Big] \Big\rangle_{i. c.}&\underset{t\rightarrow0}{=}&\frac{1}{4\pi^2}\int_{-\pi}^{\pi}dx\int_{-\pi}^{\pi}dx'\int_{0}^{+\infty}dy\,e^{-y}\;\ln\Bigg[\frac{e^{2\lambda_\alpha t+i(x'-x)}}{(1+e^{ix'})^2}\Bigg]\nonumber\\
&\underset{t\rightarrow0}{=}&2\lambda_\alpha t \; .
\end{eqnarray}
This limit corresponds to the short-time dynamics where the system remains in its initial state. Indeed in this case we have $\vec{S}(t)\approx\sqrt{N}\vec{V}_\alpha$ and thus $z\big(t,\{s_\mu(0)\}\big)=\lambda_\alpha$.

In the case where $t\rightarrow +\infty$, all terms in eq.~(\ref{eq:replicas}) are dominated by the highest eigenmode contribution. In other words the term $e^{2\lambda_N t}$ exponentially suppresses the function $f$:
\begin{eqnarray}
\frac{f(t,x,x',y)\ln[f(t,x,x',y)]}{1-f(t,x,x',y)}&\underset{t\rightarrow +\infty}{=}&f(t,x,x',y)\ln[f(t,x,x',y)]\nonumber\\
&\underset{t\rightarrow +\infty}{=}&0\;.
\end{eqnarray}
It then follows that
\begin{eqnarray}
\Big\langle \ln\Big[{\mathcal Z}_N\big(t,\{s_\mu(0)\}\big)\Big] \Big\rangle_{i. c.}&\underset{t\rightarrow +\infty}{=}&\ln\Bigg[\frac{\big\langle X^{2}\big\rangle_{i. c.}\mathcal{Y}_{\rm bulk}(t)}{2}\Bigg]+\int_{0}^{+\infty}dy\, e^{-y}\ln y\nonumber\\
&\underset{t\rightarrow +\infty}{=}&2\lambda_N t\,.
\end{eqnarray}
Here the system has fallen in the lowest eigenmode, the only stable one.

\end{document}